\documentclass[pdflatex,sn-mathphys-ay]{sn-jnl}

\makeatletter
\providecommand\@outputbox@removebskip[1]{#1} % no-op fallback for older kernels
\makeatother
\usepackage{graphicx}%
\usepackage{multirow}%
\usepackage{amsmath,amssymb,amsfonts}%
\usepackage{amsthm}%
\usepackage{mathrsfs}%
\usepackage[title]{appendix}%
\usepackage{xcolor}%
\usepackage{textcomp}%
\usepackage{booktabs}%
\usepackage{algorithm}%
\usepackage{algorithmicx}%
\usepackage{algpseudocode}%
\usepackage{listings}%
\usepackage{subcaption}
\usepackage{threeparttable}
\usepackage{tikz}
\usepackage{enumitem}
\usetikzlibrary{shapes.geometric, arrows, backgrounds, fit}
\usetikzlibrary{calc}

\tikzstyle{start} = [rectangle,           % Box shape
    rounded corners = 0.5cm,     % Rounded corners 
    minimum width=5.7cm, 
    minimum height=1.7cm, 
    text centered, 
    text width=5.0cm, 
    draw=black,
    fill=blue!20,
    font = \small,
    align = left]

\tikzstyle{start2} = [rectangle,           % Box shape
    rounded corners = 0.2cm,     % Rounded corners 
    minimum width=2.0cm, 
    minimum height=1.0cm, 
    text centered, 
    draw=black,
    fill=blue!20,
    font = \small]

\tikzstyle{NP} = [rectangle,           % Box shape
                    rounded corners = 0.5cm,     % Rounded corners 
                    minimum width=2.5cm, 
                    minimum height=1.0cm, 
                    text centered, 
                    text width=2.5cm, 
                    draw=black,
                    fill=orange!20,
                    font = \small]

\tikzstyle{process} =  [rectangle, 
                        minimum width=6.2cm, 
                        minimum height=1.5cm, 
                        text centered, 
                        text width=6.0cm, 
                        draw=black,
                        fill=gray!20,
                        font = \small,
                        align = left]

\tikzstyle{process2} =  [rectangle, 
                        minimum width=4.2cm, 
                        minimum height=1.5cm, 
                        text centered, 
                        text width=4.0cm, 
                        draw=black,
                        fill=gray!20,
                        font = \small,
                        align = left]

\tikzstyle{decision2} = [trapezium,
        trapezium stretches body,
        draw=black,
        minimum width=1cm,     % Fixed width
        minimum height=1.0cm,  % Fixed height
        trapezium left angle=120, 
        trapezium right angle=60,
        text centered,         % Center the text
        align=center,          % Center text within the node
        font = \small,
        fill=green!30]

\tikzstyle{arrow} = [thick,->,>=stealth]
\tikzstyle{line}  = [thick,-,>=stealth]

\usepackage{natbib} % For citations and bibliography
\usepackage{lmodern}
\usepackage{mathrsfs}
\usepackage{fix-cm}
\usepackage{natbib}
\usepackage[english]{babel}
\usepackage{fancyhdr}
\usepackage{multirow}
\usepackage{makecell}
\usepackage{float}
\usepackage{array}
\usepackage{setspace}
\usepackage{graphicx}%
\usepackage{multirow}%
\usepackage{amsmath,amssymb,amsfonts}%
\usepackage{amsthm}%
\usepackage{mathrsfs}%
\usepackage[title]{appendix}%
\usepackage{xcolor}%
\usepackage{textcomp}%
\usepackage{booktabs}%
\usepackage{algorithm}%
\usepackage{algorithmicx}%
\usepackage{algpseudocode}%
\usepackage{listings}%

\theoremstyle{thmstyleone}%
\theoremstyle{thmstyletwo}%

\theoremstyle{thmstylethree}%

\begin{document}

\title[Article Title]{\textbf{Efficient Multi-Objective Constrained Bayesian Optimization of Bridge Girder}}


\author*[1,2]{\fnm{Heine Havneraas} \sur{Røstum}}\email{hhr@aaj.no}
\author[3]{\fnm{Joseph} \sur{Morlier}}\email{joseph.morlier@isae-supaero.fr}
\author[1]{\fnm{Sebastien} \sur{Gros}}\email{sebastien.gros@ntnu.no}
\author[2]{\fnm{Ketil} \sur{Aas-Jakobsen}}\email{kaa@aaj.no}

\affil*[1]{\orgdiv{Engineering Cybernetics}, \orgname{Norwegian University of Science and Technology (NTNU)}, \orgaddress{\city{Trondheim}, \country{Norway}}}

\affil[2]{\orgname{Dr Ing A Aas-Jakobsen AS}, \orgaddress{\city{Oslo}, \country{Norway}}}

\affil[3]{\orgname{ICA, Université de Toulouse, ISAE-SUPAERO}, \orgaddress{\city{Toulouse}, \country{France}}}


\abstract{The buildings and construction sector is a significant source of greenhouse gas emissions, with cement production alone contributing 7~\% of global emissions and the industry as a whole accounting for approximately 37~\%. Reducing emissions by optimizing structural design can achieve significant global benefits. This article introduces an efficient multi-objective constrained Bayesian optimization approach to address this challenge. Rather than attempting to determine the full set of non-dominated solutions with arbitrary trade-offs, the approach searches for a solution matching a specified trade-off. Structural design is typically conducted using computationally expensive finite element simulations, whereas Bayesian optimization offers an efficient approach for optimizing problems that involve such high-cost simulations. The proposed method integrates proper orthogonal decomposition for dimensionality reduction of simulation results with Kriging partial least squares to enhance efficiency. Constrained expected improvement is used as an acquisition function for Bayesian optimization. The approach is demonstrated through a case study of a two-lane, three-span post-tensioned concrete bridge girder, incorporating fifteen design variables and nine constraints. A comparison with conventional design methods demonstrates the potential of this optimization approach to achieve substantial cost reductions, with savings of approximately 10\% to 15\% in financial costs and about 20\% in environmental costs for the case study, while ensuring structural integrity.}


\maketitle

\section{Introduction}
Traditionally, bridge engineers have focused on minimizing financial costs while ensuring compliance with all structural and geometric requirements. However, in recent years, the environmental impact of bridge construction has gained increasing attention. The buildings and construction sector is a major source of greenhouse gas emissions, with cement production alone accounting for 7~\% of global emissions, while the sector as a whole contributes approximately 37~\% of global emissions (\citealp{UN2023_Climate}). Reducing the emissions associated with constructions may therefore have a significant global impact. The problem is that the traditional bridge design process is heavily based on experience and manual calculations performed by highly skilled professionals. Given the numerous factors involved in bridge design, even before considering greenhouse gas emissions, finding the optimal balance between these factors has become an increasingly challenging task for human experts. As a result, many bridge designs today may be considered suboptimal.

Numerical optimization is well established (\citealp{nocedal1999numerical}), but it is often difficult to apply it directly to optimization of engineering design problems. It requires solving the system for multiple parameter values, which in engineering design consists of expensive data generation methods like numerical simulations, where gradient information is usually not directly available. This further adds complexity, making the process impractical due to excessive computational time and memory requirements. Surrogate-based optimization addresses these challenges by employing approximations of the underlying functions, known as surrogates, which are much faster to evaluate than the underlying functions during the numerical optimization process (\citealp{martins2021engineering}). This approach has been shown to greatly reduce computational time in bridge design (\citealp{penades2019accelerated}). A similar approach is Bayesian optimization, which builds probabilistic surrogates, where the probabilistic element is leveraged to accelerate the optimization process (\citealp{frazier2018bayesian}). The probabilistic surrogate is typically constructed using Gaussian Process Regression (GPR). Bayesian optimization has been successfully applied to various engineering design problems; materials (\citealp{frazier2016bayesian}), civil structures (\citealp{mathern2021multi}), aerospace (\citealp{tfaily2024bayesian, saves2024gaussian}), and bridge design (\citealp{rostum2025constrained}). It ensures fast convergence in the optimization process for problems that typically involve less than 20 decision variables (\citealp{frazier2018bayesian}).

A challenge in applying GPR is its poor scalability with dataset size, as the training time increases cubically with the number of data points. For high-dimensional simulations, such as those obtained from \emph{Finite Element} (FE) analyses, this can make training computationally intractable if attempting to approximate the full high-dimensional output using a single surrogate model. To mitigate this issue, \emph{Reduced Order Modeling} (ROM) has been widely adopted (\citealp{lucia2004reduced, peherstorfer2015dynamic, chiplunkar2018gaussian, drakoulas2024physics}). ROM replaces the full-order model with a lower-dimensional approximation, significantly reducing both memory requirements and computational costs. A widely used technique for ROM is the \emph{Proper Orthogonal Decomposition} (POD) (\citealp{liang2002proper, amsallem2012stabilization, paul2015adaptive, guo2018reduced, zhao2021reduced, rocha2023reduced}). \cite{guo2018reduced} used POD with GPR to predict unobserved design variable combinations of the full solution field of non-linear FE analyses. Their approach showed good performance in both accuracy and computational efficiency. A challenge with surrogate models is that the number of design variables, or input parameters, affects its accuracy and training time. To address this, \cite{bouhlel2016improving} proposed a method called \emph{Kriging Partial Least Squares} (KPLS) where \emph{Partial Least Squares} (PLS) is applied for the construction of the GPR. Their results showed that with the same number of training points, they could significantly reduce training time while achieving similar or improved accuracy compared to standard GPR when applied to problems with a moderate to high number of design variables. 

Using numerical optimization in the design of new structures can be an effective way to reduce global emissions. However, for this approach to be practical, it must be efficient, both in terms of training time and the number of iterations required for the optimizer to converge. Reducing the number of iterations reduces the number of potentially time-consuming numerical simulations, significantly impacting the overall duration of the optimization process. To further enhance the appeal of this approach, the optimization process should consider both the environmental impact and the monetary cost of the structure. By striking a balance between these objectives, it is possible to achieve trade-offs that reduce both, offering improvements over traditional design methods also in the monetary sense. To achieve this, this article presents a method for multi-objective constrained Bayesian optimization using a predefined trade-off function. 

Early extensions of Bayesian optimization to multi-objective problems were based on the Efficient Global Optimization (EGO) framework \citep{jones1998efficient}. One of the first was \emph{ParEGO} \citep{knowles2005parego}, which employs randomly generated weight vectors in a Tchebycheff scalarization to enable the use of expected improvement within a single-objective EGO loop. Simultaneously, \cite{jeong2005efficient} applied EGO directly to multi-objective aerodynamic design by formulating expected improvements for each objective and identifying non-dominated solutions. In recent years, the \emph{Expected Hypervolume Improvement} (EHVI) criterion has become increasingly popular for guiding Bayesian multi-objective optimization \citep{yang2019efficient, mathern2021multi}.

However, while these approaches are effective in exploring the Pareto front, they do not directly address the practical scenario in which only one final design will ultimately be implemented. To this end, our method targets a single solution aligned with a predefined trade-off, thereby requiring fewer iterations and number of numerical simulations than traditional multi-objective optimization. The approach is conceptually similar to the work of \citet{jeong2005efficient}, who selected three representative points on the Pareto front approximation based on the objectives’ expected improvements (the extremities and the midpoint). We extend this idea by incorporating constrained expected improvement and selecting the point that maximizes the aggregated improvement over the current best feasible design. Furthermore, the proposed method utilizes the well-known POD to reduce the high-dimensional space from numerical simulations. This is combined with KPLS to obtain a computationally efficient approach to optimize engineering designs based on simulations.

The approach is applied to a case study of a two-lane, three-span post-tensioned concrete bridge girder. This case study considers fifteen design variables and nine high-dimensional constraints derived from simulations, making it a realistic and relevant example for demonstrating the method’s applicability to practical engineering problems. The objectives are to minimize the monetary and environmental cost of the structure, while satisfying all constraints. To assess the effectiveness of using a predefined trade-off function for the objectives, as opposed to the traditional arbitrary trade-off, the approach is compared to the constrained multi-objective Bayesian optimization scheme \emph{constrained expected hypervolume improvement} (\citealp{daulton2020differentiable}). Results are benchmarked against a conventional bridge-girder design approach, which is based on the authors' experience in the bridge engineering industry. This comparison highlights the potential benefits of applying the constrained multi-objective optimization approach, demonstrating its ability to improve both efficiency and performance in bridge design. 

\section{Methodology}\label{Ch:Methodology}
This section presents the constrained multi-objective optimization methodology which was applied to optimize the design of the bridge girder in the case study. Though the method is here applied to a bridge girder, the methodology is applicable to any problem that involves;
\begin{itemize} [itemsep=6pt, topsep=0pt]
    \item \emph{Expensive Data Generation (e.g., simulation results)}\\
    The high computational cost of each simulation limits the available data. Reducing the number of required simulations is crucial due to time and memory constraints.

    \item \emph{Small to Moderate Number of Decision Variables}\\
    Bayesian optimization struggles to scale with a large number of decision variables due to the \emph{curse of dimensionality}. A higher-dimensional space demands more data, which GPR handles inefficiently. KPLS improves accuracy in moderate-dimensional problems with limited data, enhancing the effectiveness of the proposed approach.
\end{itemize}
These conditions are common in many engineering design problems. The section starts with a description of the initial sampling using \emph{Latin Hypercube Sampling}, followed by a description of POD, which is used to create a low-dimensional approximation of the high-dimensional, also known as functional, simulation outputs from the initial observations. To predict unobserved variable combinations, GPR is used to interpolate the low-dimensional POD bases. Additionally, PLS is employed to reduce the number of GPR kernel parameters, with this reduction of kernel parameters referred to as KPLS. As GPR is a probabilistic regressor, the probabilistic element of it may be exploited in a Bayesian optimization scheme using a Bayesian acquisition function. If the constraints are also approximated using GPR, they can be included to formulate a constrained Bayesian optimization scheme, which is done in this article. Since multiple objective functions are involved, each one is defined using a constrained Bayesian acquisition function and incorporated into a \emph{multi-objective optimization} (MOO) framework, which will be detailed at the end of this section. Rather than approximating the entire set of non-dominated solutions, this MOO approach focuses on efficiently identifying a single solution that best satisfies a predefined trade-off between the objectives, aiming to do so in as few iterations as possible.

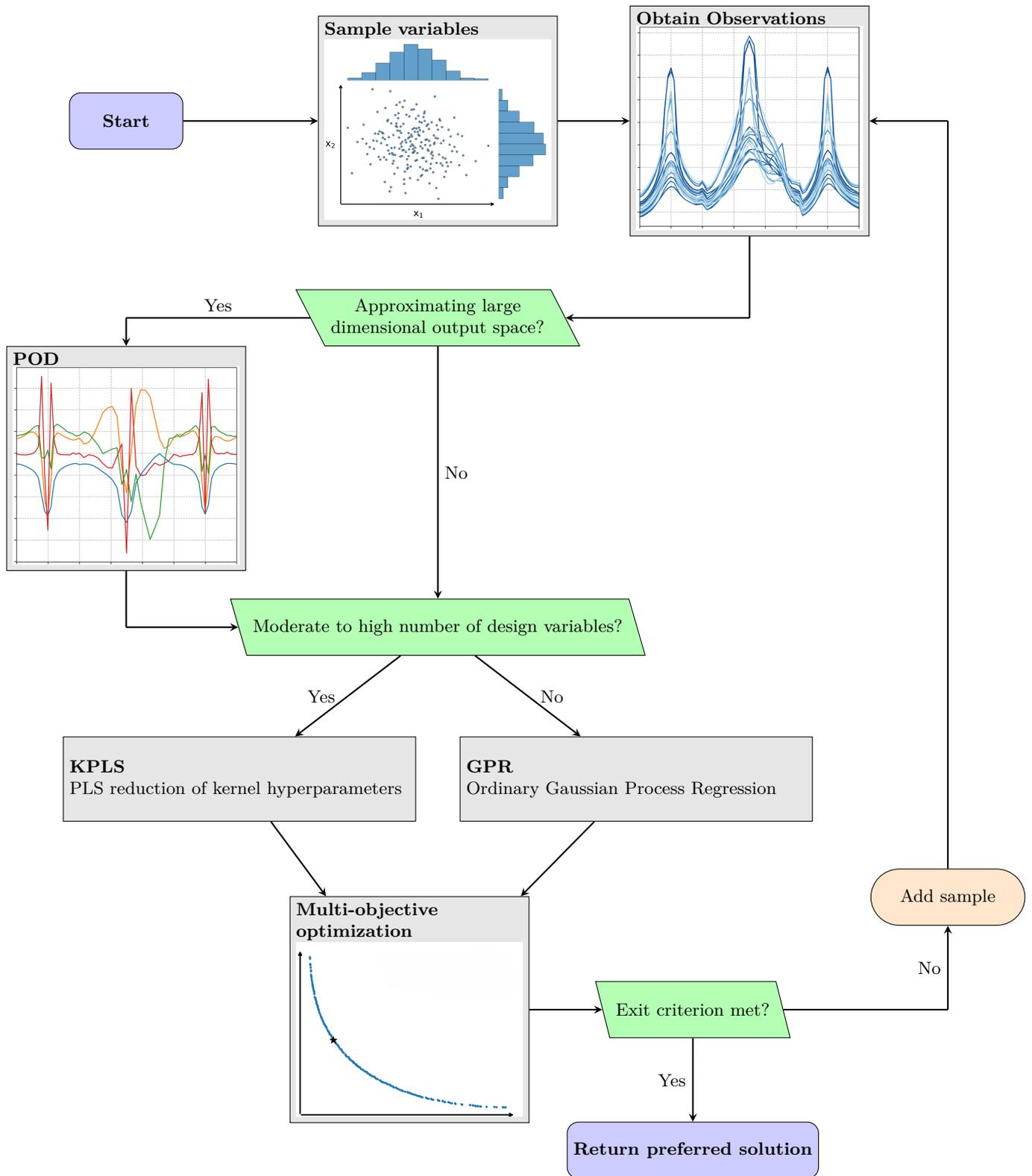
\begin{figure}[H]
    \centering
        \begin{tikzpicture}[node distance=2.5cm]
        \small
        \node (ID) [start2] {\textbf{Start} 
                };
        \node (LHS) [process2, right of = ID, xshift = 3.0cm] {\textbf{Sample variables} \\
        \includegraphics[width=4.0cm]{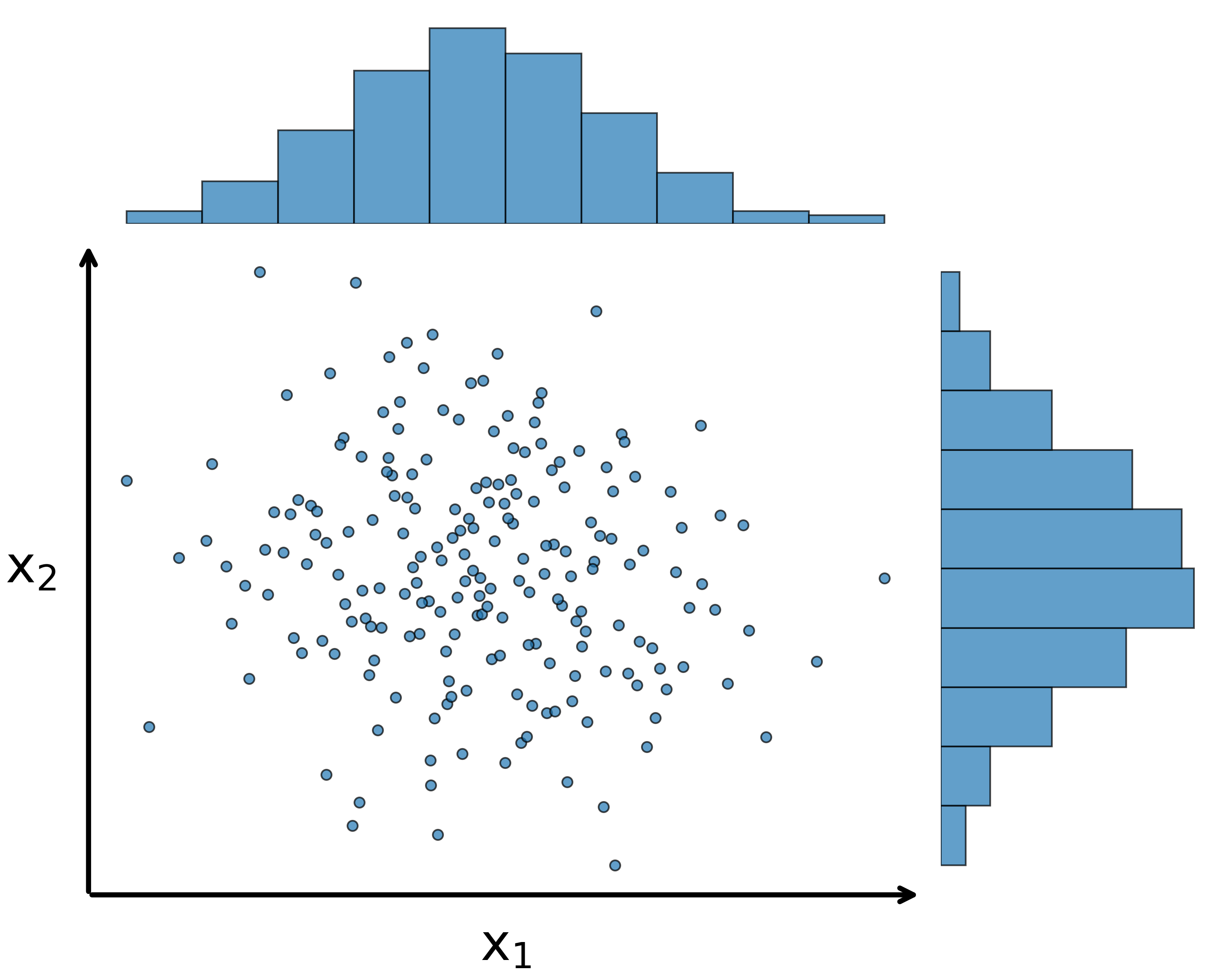}
        };
        \node (Observations) [process2, right of = LHS, xshift = 3.0cm] {\textbf{Obtain Observations} \\
        \includegraphics[width=4.0cm]{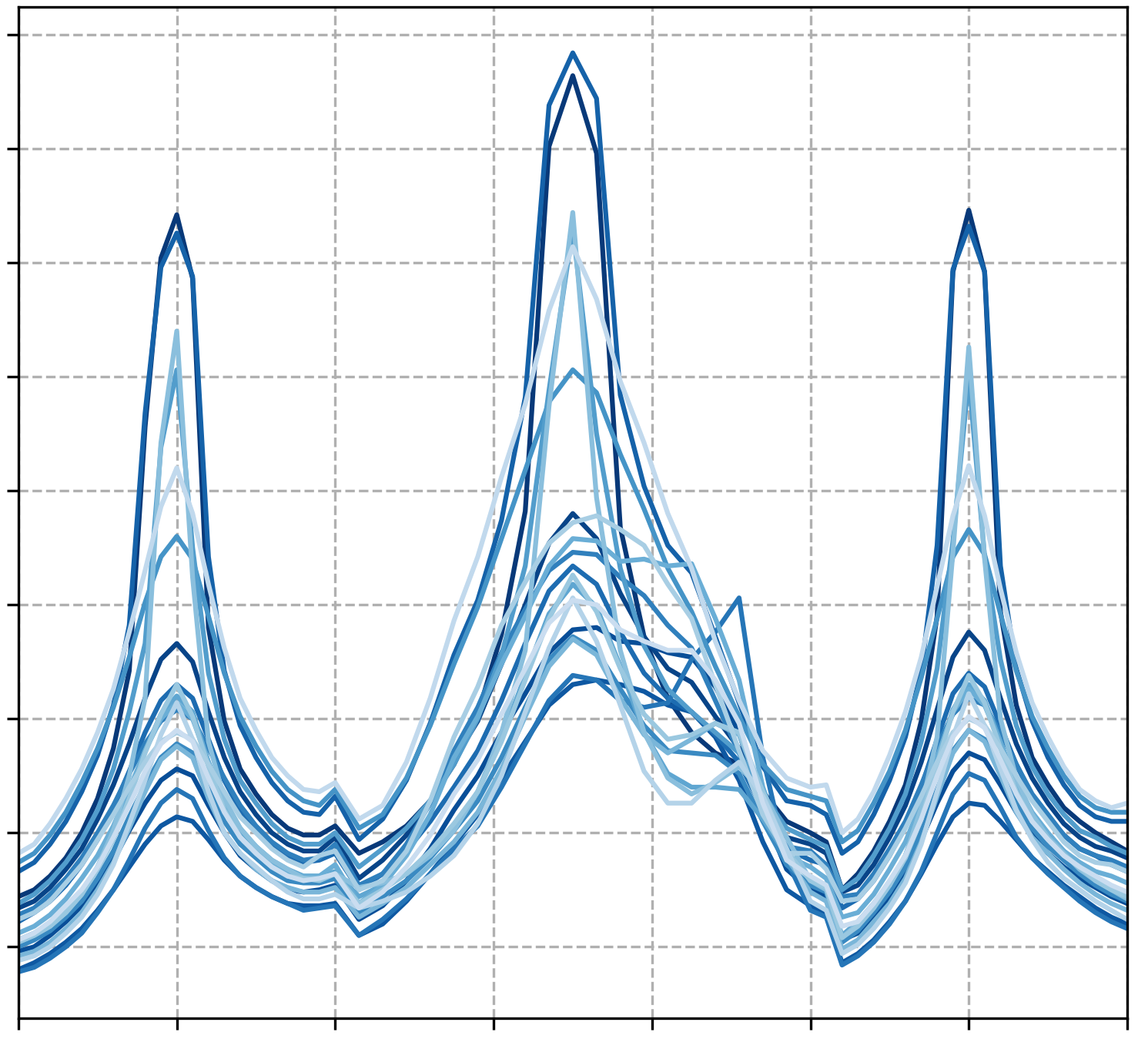}};
        \node (HighDim) [decision2, below of = LHS, yshift = -1.0cm] {Approximating large \\dimensional output space?};
        \node (POD) [process2, below of = HighDim, xshift = -5.5cm] {\textbf{POD}\\
        \includegraphics[width=4.0cm]{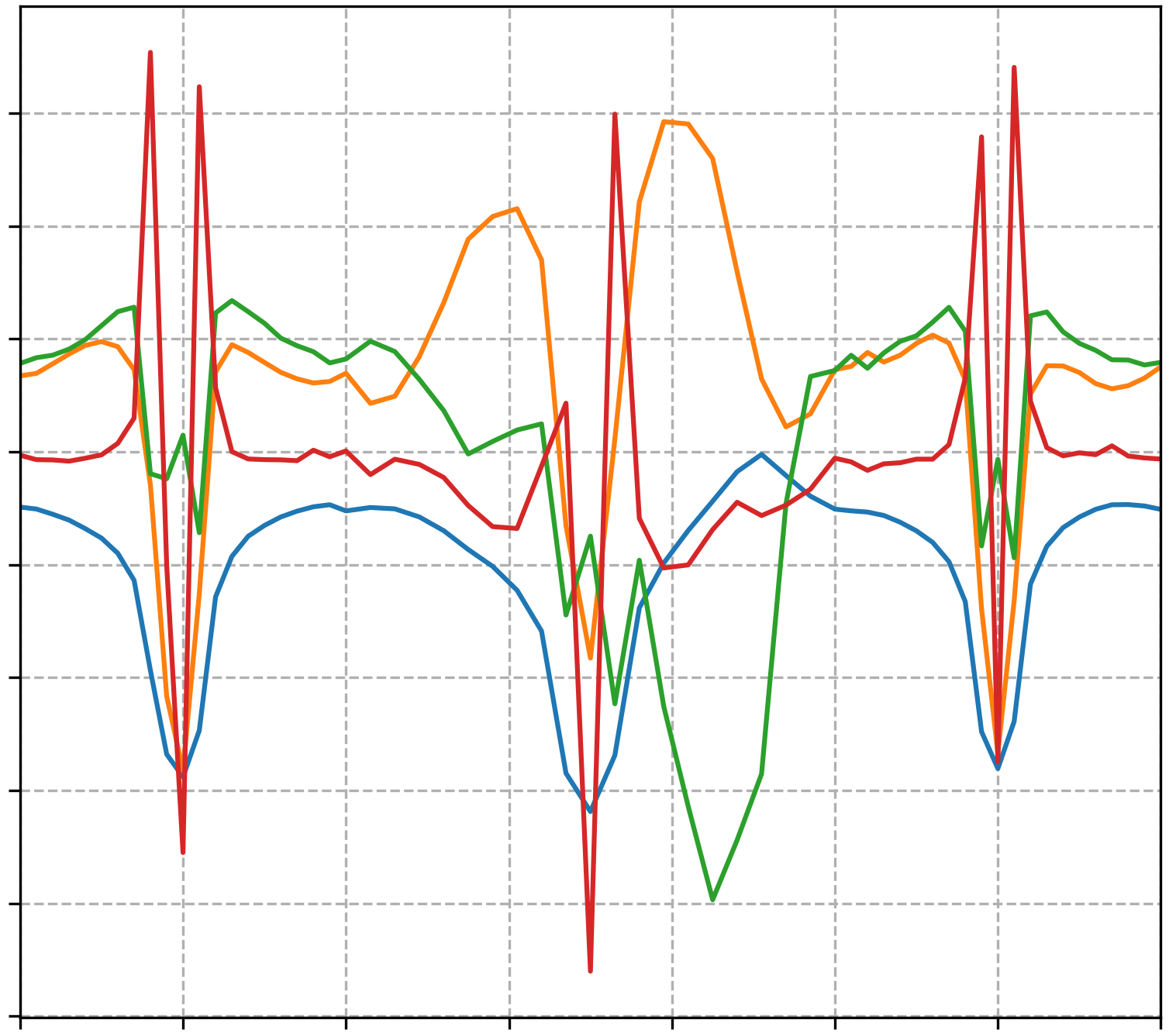}
        };
        \node (HighVar) [decision2, below of = HighDim, yshift = -3.0cm] {Moderate to high number of design variables?};
        \node (KPLS) [process, below of = HighVar, xshift = -3.5cm, yshift = -0.2cm] {\textbf{KPLS}\\ PLS reduction of kernel hyperparameters};
        \node (GPR) [process, below of = HighVar, xshift = 3.5cm, yshift = -0.2cm] {\textbf{GPR}\\ Ordinary Gaussian Process Regression};
        \node (MOO) [process2, below of = KPLS, xshift = 3.0cm, yshift = -1.6cm] {\textbf{Multi-objective \\ optimization}\\
        \includegraphics[width=4.0cm]{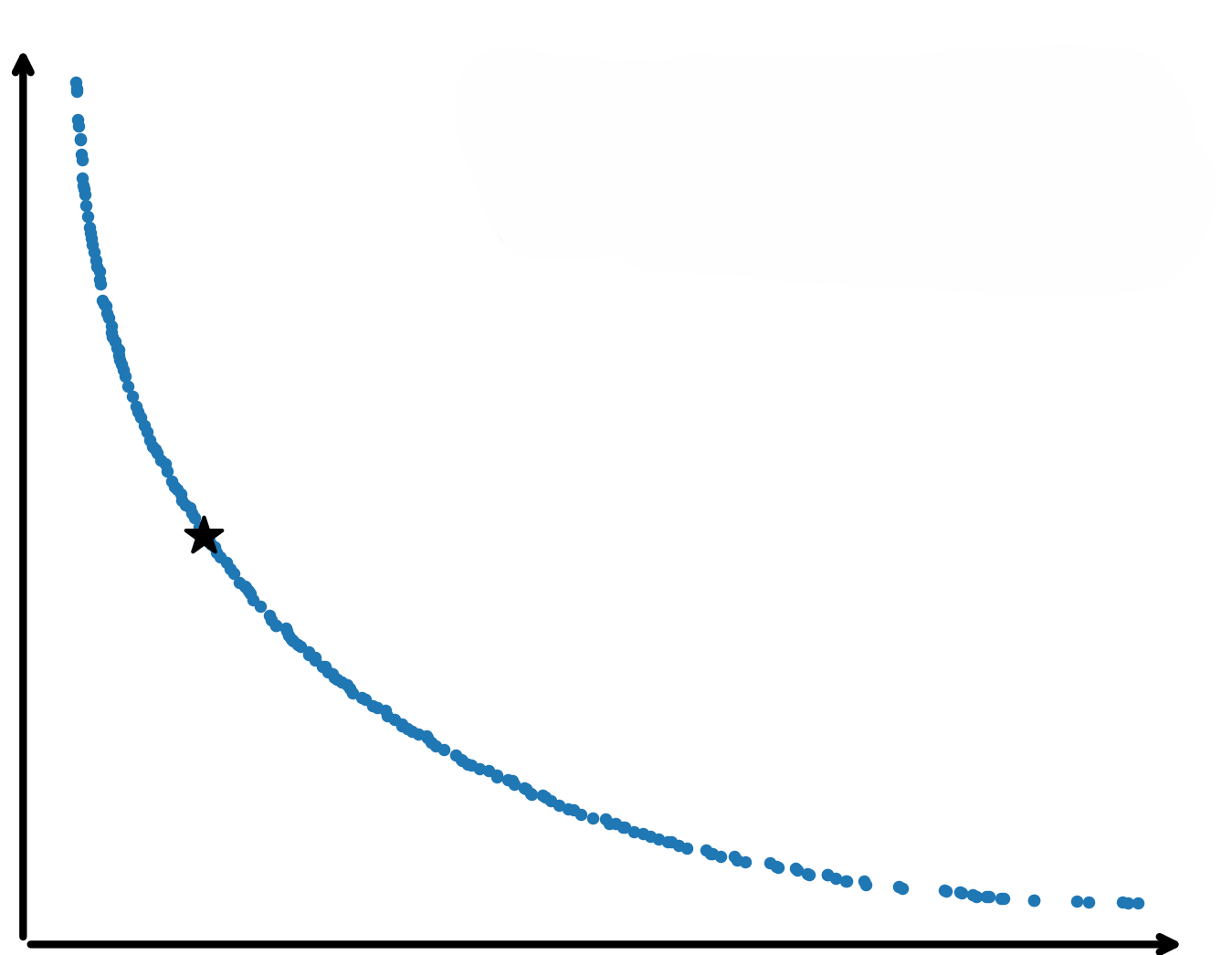}};
        \node (Res) [decision2, right of = MOO, xshift = 2.5cm] {Exit criterion met?};
        \node (x_opt) [start2, below of = Res, yshift = -0.0cm] {\textbf{Return preferred solution}};
        \node (Np1) [NP, right of = MOO, xshift = 7.0cm, yshift = 2.0cm] {Add sample};

        \draw [arrow] (ID) -- (LHS);
        \draw [arrow] (LHS) -- (Observations);
        \draw [line] (Observations.south) -- ($(Observations.south |- HighDim.east)$);
        \draw [arrow] ($(Observations.south |- HighDim.east)$) -- (HighDim.east);
        \draw [arrow] (HighDim) -- node[anchor=west] {No} (HighVar);
        \draw [line] (HighDim.west) -- node[anchor=south] {Yes} ($(POD.north |- HighDim.west)$);
        \draw [arrow] ($(POD.north |- HighDim.west)$) -- (POD.north);
        \draw [line] (POD.south) -- ($(POD.south |- HighVar.west)$);
        \draw [arrow] ($(POD.south |- HighVar.west)$) -- (HighVar.west);
        \draw [arrow] (HighVar) -- node[anchor=west, xshift = 0.1cm] {No} (GPR);
        \draw [arrow] (HighVar) -- node[anchor=east, xshift = -0.1cm] {Yes} (KPLS);
        \draw [arrow] (GPR) -- (MOO);
        \draw [arrow] (KPLS) -- (MOO);
        \draw [arrow] (MOO) -- (Res);
        \draw [arrow] (Res) -- node[anchor=east] {Yes} (x_opt);
        \draw [line] (Res.east) -- ($(Np1.south |- Res.east)$);
        \draw [arrow] ($(Np1.south |- Res.east)$) -- node[anchor=east] {No} (Np1.south);
        \draw [line] (Np1.north) -- ($(Np1.north |- Observations.east)$);
        \draw [arrow] ($(Np1.north |- Observations.east)$) -- (Observations.east);
        
        \end{tikzpicture}
    \caption{Illustration of the proposed efficient multi-objective Bayesian optimization process}
    \label{fig:OptimizationProcess}
\end{figure}

The MOO process is illustrated in Figure \ref{fig:OptimizationProcess}. The figure illustrates that the POD and KPLS can be used together or separately. The POD is applied when one is approximating the full solution field of a large dimensional output space, which could be the results from \emph{Finite Element Method} (FEM) or \emph{Computational Fluid Dynamics} (CFD) simulations. If a moderate to high number of design variables are considered, KPLS should be used as a surrogate model instead of ordinary GPR. A multi-objective optimization is subsequently conducted, using a preferred trade-off between the objectives, which will be further described at the end of this section. If the exit criterion is satisfied, the best observed solution according to the specified trade-off is returned; otherwise, the next candidate solution is incorporated as a new sample and the process is repeated.

In this article, we use bold uppercase letters to denote matrices, bold lowercase letters
to denote vectors, and plain lowercase letters to denote a scalar or variable. 

\subsection{Initial Sampling}
\begin{figure}[H]
    \centering
    \includegraphics[width=0.40\textwidth]{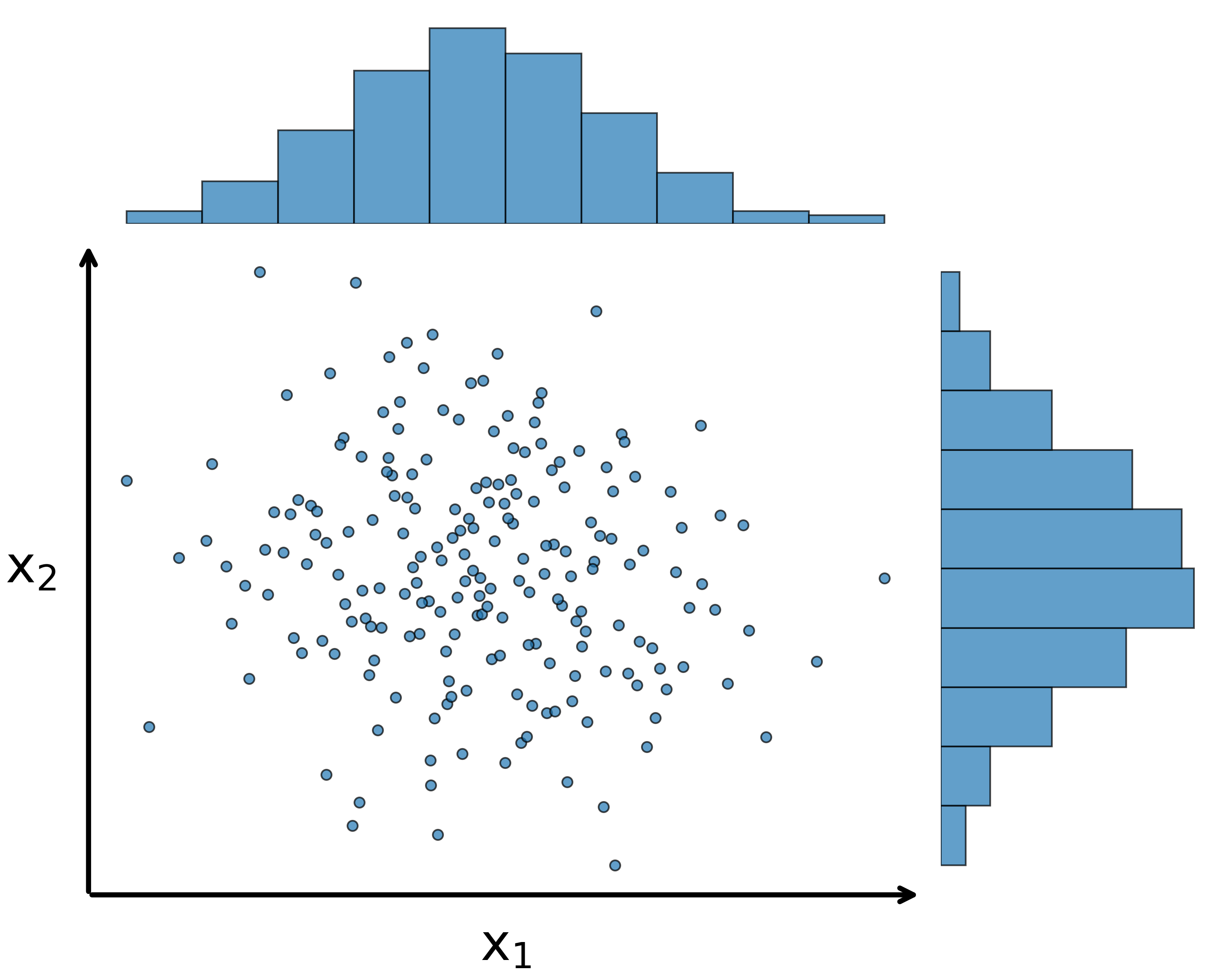}
    \caption{Illustration of two-variable latin hypercube sampling with distribution projected to axes}
    \label{fig:LHS_normal}
\end{figure}

The initial sampling method, also sometimes known as Design of Experiments (DoE), is the process of determining the initial combinations of design variables, $\textbf{X} = \left\{ \textbf{x}_j \right\}_{j=1}^{N_s}$, where $\textbf{x}_j \in \mathbb{R}^{d}$, from which to acquire observations. An observation could be the results of FEM analyses or from other calculations using a combination of design variables $\textbf{x}_j$. $N_s$ is the number of combinations of design variables to sample, or the sampling budget, and $d$ is the number of design variables. From the initial combination of design variables, an observation set $\mathcal{D} = \left\{ (\textbf{x}_j, \textbf{y}(\textbf{x}_j)) \right\}_{j=1}^{N_s}$ is acquired. A popular sampling strategy is the \emph{Latin Hypercube Sampling} (LHS) strategy (\citealp{LiteratureReview_MetamodelStructures}). LHS was first described by \cite{artLHS}. The LHS strategy implemented in this article is to generate a LHS with a normalized distribution in the direction of each variable. The technique is similar to that described in the book by \cite{martins2021engineering} for LHS with normalized distribution. An illustration of a two-variable LHS with distributions projected to the axes is shown in Figure \ref{fig:LHS_normal}.
A normalized distribution is chosen to ensure that not multiple design variables are in the variable bounds' extremities for the same variable combination, as this may result in analytical problems during the FEM analyses. This depends on the type of analysis and software that is used for the simulations. However, it should be noted that in high-dimensional spaces, much of the volume of the design domain lies near the boundaries. Under-sampling these regions may lead to over-exploration by the acquisition function in later iterations due to the high associated uncertainty. A uniform distribution may typically be more applicable to most situations and ensure that more information is obtained from the initial sampling. 

\subsection{Proper Orthogonal Decomposition (POD)}\label{Ch.POD}
Approximating the entire output space from FEM simulations, such as e.g. displacements for the entire structure, is typically computationally demanding using a single GPR due to the high dimensionality of the results. POD is a data analysis method with the purpose of obtaining a low-dimensional approximation of a high-dimensional process (\citealp{chatterjee2000introduction}). If each result for the entire output space for different combinations of design variables is stored in a vector $\textbf{y}(\textbf{x}) \in \mathbb{R}^{N_h}$, where $N_h$ is the number of datapoints of the high-dimensional output. The results from the initial sampling can be  stored in a matrix $\textbf{Y} \in \mathbb{R}^{N_h \times N_s}$;
\begin{equation}  \label{Eq.Y}
    \textbf{Y} = \begin{bmatrix} 
        \text{y}_{1,1} & \cdots & \text{y}_{1,N_s}\\ 
        \vdots & \ddots & \vdots\\ 
        \text{y}_{N_h,1} & \cdots & \text{y}_{N_h,N_s}
    \end{bmatrix}
\end{equation}
$\textbf{y}$ and $\textbf{Y}$ are here used as general terms to denote observed responses, which could be constraints or objectives. In the context of POD the $\textbf{Y}$ matrix is often known as a \emph{snapshot matrix}, where the number of simulations $N_s$ is the number of \emph{snapshots}. We assume that all simulations share the same underlying mesh. If they do not, their outputs must first be interpolated or projected onto a common reference grid to maintain consistency. POD is commonly achieved using the \emph{Singular Value Decomposition} (SVD) (\citealp{guo2018reduced}). Applying SVD to the matrix $\textbf{Y}$ yields:
\begin{equation}
    \textbf{Y} = \textbf{U} \boldsymbol{\Sigma} \textbf{V}^\top
\end{equation}
Here $\textbf{U} \in \mathbb{R}^{N_h \times N_s}$ and $\textbf{V} \in \mathbb{R}^{N_s \times N_s}$ are orthogonal matrices, and  $\boldsymbol{\Sigma} = \text{diag} \left\{ \sigma_1, \sigma_2, ..., \sigma_{N_s} \right\}$ are the singular values (eigenvalues) in descending order. A low dimensional truncated POD basis, $\textbf{L} \in \mathbb{R}^{N_h \times N_{POD}}$, which is derived by preserving the first $N_{POD}$ modes of $\textbf{U}$ that corresponds to the largest singular values, can be used to calculate the \emph{projection coefficients} $\textbf{A} \in \mathbb{R}^{N_{POD} \times N_{s}}$, also sometimes known as the \emph{amplitude matrix} (\citealp{buljak2011proper, de2021evaluation}), using the relationship:
\begin{equation}
    \textbf{A} = \textbf{L}^\top \cdot \textbf{Y}
\end{equation}
A common method of selecting the number of POD modes to include is by evaluating the projection error $\epsilon_{POD}$ of including the first $N_{POD}$ left singular vectors of $\textbf{Y}$ (\citealp{eckart1936approximation, guo2018reduced, brunton2022data}). Thus $N_{POD}$ can be defined as the minimum integer such that:
\begin{equation} \label{Eq: Number of POD modes}
    \frac{\sum_{i=1}^{N_{POD}} \sigma_i^2}{\sum_{i=1}^{N_s} \sigma_i^2} > 1 - {\epsilon_{\text{POD}}}^2
\end{equation}
In this article an ${\epsilon_{\text{POD}}}^2$ of 0.01 is used for selecting the number of POD modes, $N_{POD}$. Due to the orthogonality of $\textbf{U}$, an approximation of the snapshot matrix \textbf{Y} can be made from the low dimensional POD basis $\textbf{L}$ and the projection coefficients $\textbf{A}$ as:
\begin{equation}
    \textbf{Y} \approx \textbf{L} \cdot \textbf{A}
\end{equation}

Using the training set $\mathcal{D} = \left\{ (\textbf{X}, \textbf{A}_i) \right\}_{i=1}^{N_{POD}}$, one regression model $\hat{\textbf{A}}_i(\textbf{x})$ is created to approximate each of the $N_{POD}$ projection coefficients in $\textbf{A} \in \mathbb{R}^{N_{POD} \times N_{s}}$, resulting in the predicted projection coefficients:
\begin{equation}
    \hat{\textbf{A}}(\textbf{x}) = \left[\hat{\text{A}}_1(\textbf{x}), \hat{\text{A}}_2(\textbf{x}), \ldots, \hat{\text{A}}_{N_{POD}}(\textbf{x}) \right]^\top
\end{equation}
Approximations of the full set solution $\hat{\textbf{Y}}$ can be retrieved for new variable combinations $\textbf{x}$ using the equation:
\begin{equation}
    \hat{\textbf{Y}}(\textbf{x}) = \textbf{L} \cdot \hat{\textbf{A}}(\textbf{x})
\end{equation}
In this article, Gaussian process regression (GPR) is applied as the regression model, which will be further described in the following section.

\subsection{Gaussian Process Regression (GPR)}
A Gaussian process is a collection of random variables, any finite number of which have a joint Gaussian distribution (\citealp{Book_GPForML}). A real process, $h(\textbf{x})$ can be formulated completely as a Gaussian process by its mean function, $\mu(\textbf{x})$, and covariance function $k(\textbf{x},\textbf{x}')$, with the Gaussian process formulated as:
\begin{equation}
    h(\textbf{x}) \sim \mathcal{GP}(\mu(\textbf{x}), k(\textbf{x},\textbf{x}'))
\end{equation}

The mean is often assumed to be zero for notational simplicity, as will be done in this article, and this can be approximated by \emph{mean centering} the distribution. The covariance function, $ k(\textbf{x},\textbf{x}')$, typically determines the GP's smoothness, and characteristic length-scale. In this article, the Matérn 5/2 kernel is used, which is formulated as (\citealp{Book_GPForML}):
\begin{equation}
    k(\textbf{x},\textbf{x}') = \sigma_f^2\prod_{i=1}^d\left(1+\sqrt{5}r_i+\frac{5}{3}{r_i}^2\right)\exp\left(-\sqrt{5}r_i\right)
\end{equation}
where $d$ is the number of design variables, $\sigma_f^2$ is the signal variance, and $r_i$ is a distance metric for the $i^{\text{th}}$ design variable component. In this article a distance metric with an anisotropic length-scale $\theta_i$ is used:
\begin{equation}
    r_i = \theta_i\left| \text{x}_i-\text{x}'_i\right|
\end{equation}
When a set of $N_s$ observations are available, the observations $\mathcal{D} = \{(\textbf{x}_j, \text{y}_j)\}_{j=1}^{N_s}$, can be used to in order to infer a joint Gaussian posterior distribution for unobserved data, $\textbf{y}_*(\textbf{X}_*)$:
\begin{equation}
    \textbf{y}_*|\textbf{X}_*,\mathcal{D} \sim \mathcal{N}(\hat{\mu}(\textbf{X}_*), \hat{\sigma}^2(\textbf{X}_*))
\end{equation}
Where the posterior mean, $\hat{\mu}(\textbf{X}_*)$, and variance, $\hat{\sigma}^2(\textbf{X}_*)$, is calculated as:
\begin{subequations}\label{Eq:GPRpred}
\begin{align}  
    \hat{\mu}(\textbf{X}_*) = & K(\textbf{X}_*,\textbf{X})[K(\textbf{X},\textbf{X})+\sigma_n^2\textbf{I}]^{-1} \textbf{y} \label{Eq:GPRmean} \\
    \hat{\sigma}^2(\textbf{X}_*) = & K(\textbf{X}_*,\textbf{X}_*)-K(\textbf{X}_*,\textbf{X})[K(\textbf{X},\textbf{X}) + \sigma_n^2\textbf{I}]^{-1}K(\textbf{X},\textbf{X}_*) \label{Eq:GPRvar}
\end{align}
\end{subequations}

For compactness, $K(\textbf{X},\textbf{X})$ will be abbreviated as $\textbf{K}_\text{ff}$, and $K(\textbf{X}_*,\textbf{X}_*)$ as $\textbf{K}_{**}$, and so forth. The $\sigma_n^2 \textbf{I}$ term is the deviation between the observations and the regressor. $\sigma_n^2$ is the deviation variance, which is often referred to as the \emph{noise variance}, and $\textbf{I}\in \mathbb{R}^{N_s \times N_s}$ is an identity matrix. Though the results from the FEM simulations are noise free, $\sigma_n^2$ is included as a regularization term.

The \emph{marginal likelihood} is commonly used to determine the model hyperparameters $\boldsymbol{\beta} = \left\{\sigma_f^2, \theta_1, \ldots, \theta_{d}, \sigma_n^2\right\}$, which consist of the noise variance, signal variance, and the parameters associated to the covariance function. The \emph{logarithmic marginal likelihood} is formulated as:
\begin{equation} \label{Eq:marglikelihood}
\log  p(\textbf{y}\mid \textbf{X}) = -\frac{1}{2}\log\det(\textbf{K}_\text{ff} +\sigma_n^2\textbf{I})-\frac{1}{2} \textbf{y}^\top (\textbf{K}_\text{ff}+\sigma_n^2\textbf{I})^{-1} \textbf{y} -\frac{n}{2}\log(2\pi)
\end{equation}
The optimal hyperparameters $\boldsymbol{\beta}_\text{opt}$ are then typically obtained by the maximization problem:
\begin{equation}
    \boldsymbol{\beta}_\text{opt} = \arg\max_{\boldsymbol{\beta}} \log  p(\textbf{y}\mid \textbf{X})
\end{equation}

Which is also used in this article. The first term in Eq. \eqref{Eq:marglikelihood} penalizes high complexity, in the sense that large values of $\textbf{K}$ is discouraged in the covariance function, as it results in non-smooth behavior. The second term ensures a good data fit, and the last term is a normalization constant. Optimizing the hyperparameters using the marginal likelihood thus ensures a good fit for the observed values, and reduces the risk of overfitting by penalizing complex models. 

\subsection{Kriging partial least squares (KPLS)}
\cite{bouhlel2016improving} showed that weights derived from \emph{Partial Least Squares (PLS)} could be used to efficiently reduce the number of kernel hyperparameters associated to the design variables from $\left\{\theta_i \right\}_{i=1}^d$ to $\left\{\theta_l \right\}_{l=1}^h$, where $h < d$. They called this method KPLS, and showed that the method resulted in a significant reduction in the CPU training time, and that the method in general provided comparable or better predictive metrics for problems involving a medium to high number of hyperparameters. The method relies on PLS, which is a method for determining an optimum linear relationship between inputs $\textbf{X}$ and outputs $\textbf{y}$ (\citealp{helland1988structure}). The number of components, $h$, to include in the parameter reduction is chosen a priori, but a low number of weights is typically selected. By applying PLS with $h$ number of components to observed inputs $\textbf{X} \in \mathbb{R}^{N_s \times d}$ and outputs $\textbf{y}(\textbf{X}) \in \mathbb{R}^{N_s}$, one obtains the PLS weights $\textbf{W} \in \mathbb{R}^{d \times h}$ and loadings $\textbf{P} \in \mathbb{R}^{d \times h}$. A new projected coordinate system, $\textbf{T} \in \mathbb{R}^{N_s \times h}$, obtained by rotating the original system, can be formulated as (\citealp{bouhlel2016improving}):
\begin{equation}
    \textbf{T} = \textbf{X}\textbf{W}_*
\end{equation}
Where $\textbf{W}_*$ is known as the projection or \emph{rotation} matrix, and can be calculated using the relationship:
\begin{equation}
    \textbf{W}_*=\textbf{W}\left[\textbf{P}^\top\textbf{W}\right]^{-1}
\end{equation}
The rotation matrix $\textbf{W}_*=\left[ \textbf{w}_*^{(1)},\ldots, \textbf{w}_*^{(h)}\right]$, is used to reduce the number of hyperparameters associated to the length-scales $\boldsymbol{\theta}$ in the kernel from $d$ to $h$. The KPLS version of the Matérn 5/2 kernel is given in Eq. \eqref{Eq.KPLS_Matern}.

\begin{equation}\label{Eq.KPLS_Matern}
    k(\textbf{x},\textbf{x}') = \sigma_f^2\prod_{l=1}^h\prod_{i=1}^d\left[1+\sqrt{5}r_i^{(l)}+\frac{5}{3}\left({r_i^{(l)}}\right)^2\right]\exp\left(-\sqrt{5}r_i^{(l)}\right)
\end{equation}
Here, the rotation coefficient $w_{*i}^{(l)}$ is included in the distance metric $r_i^{(l)}$ as given in Eq. \eqref{Eq.KPLS_DistanceMetric}.
\begin{equation}\label{Eq.KPLS_DistanceMetric}
    r_i^{(l)} = \theta_l\left| w_{*i}^{(l)}\left(\text{x}_i-\text{x}'_i\right)\right|
\end{equation}

\subsection{Optimization process}\label{Ch.OptProc}
A typical multi-objective optimization problem of determining the optimum design variables $\textbf{x}^*$ subjected to inequality constraints can mathematically be formulated as: 
\begin{subequations} \label{Eq. MOO}
\begin{align}
\textbf{x}^* = \underset{\textbf{x} \in \textbf{X}}{\text{arg min}} &\; \left( f_1(\textbf{x}),f_2(\textbf{x}),\ldots,f_{n_f}(\textbf{x}) \right)\\
\mathrm{s.t}\quad& g_i(\textbf{x}) \leq \lambda_i, \quad i = 1, 2,\ldots,n_c
\end{align}
\end{subequations}
where $\boldsymbol{f}(\textbf{x}) = \left( f_1(\textbf{x}),f_2(\textbf{x}),\ldots,f_{n_f}(\textbf{x}) \right)$ are the objective functions, subjected to $n_c$ number of inequality constraints expressed by the $i^{th}$ constraint function as $g_i(\textbf{x})$. The goal is often to identify the set of feasible solutions for which no other feasible solution dominates them. A solution $\textbf{x}^* \in \textbf{X}$ is non-dominated if there exists no other feasible solution $\textbf{x}' \in \textbf{X}$ such that $\boldsymbol{f}(\textbf{x}') \prec \boldsymbol{f}(\textbf{x}^*)$. The corresponding set of feasible non-dominated solutions is known as the Pareto front approximation $\mathcal{P} = \left\{\boldsymbol{f}^{(1)},\ldots, \boldsymbol{f}^{(n_\mathcal{P})} \right\} \subset \mathbb{R}^{n_f}$, where $n_\mathcal{P}$ is the number of non-dominated solutions. If the objective and constraint functions are approximated by probabilistic \emph{surrogate models}, such as GPR, the probabilistic element can be exploited in a \emph{Bayesian} optimization framework. In the Bayesian optimization framework, the original constrained multi-objective problem defined in Eq.~\eqref{Eq. MOO} is reformulated using an acquisition function that guides the selection of the next design (or sampling) point. This acquisition function typically transforms the multi-objective minimization problem into a scalar maximization problem. Whether the goal is to approximate the entire Pareto front or to target a specific trade-off solution, the optimization process is generally performed one point at a time.

We first present a popular acquisition function for single-objective constrained Bayesian optimization. This is followed by extensions of it for multi-objective constrained Bayesian optimization. In the framework of multi-objective constrained Bayesian optimization, we consider two distinct methodologies:
\begin{enumerate}
    \item \textbf{The method proposed in this work}, which targets a specific solution or trade-off among the objective functions. The goal is to identify a single desirable design as efficiently as possible, rather than deriving the entire Pareto front.
    \item \textbf{The constrained expected hypervolume improvement approach}, which represents a standard, off-the-shelf multi-objective Bayesian optimization method. It is designed to explore the objective space broadly and incrementally improve the entire Pareto front. This serves as a reference for comparison.
\end{enumerate}

The purpose of the proposed approach is to rapidly converge to a single balanced trade-off point by maximizing an acquisition function based on normalized objective values. Since only one structural design will ultimately be implemented, our method avoids the computational overhead of constructing a full Pareto front and instead prioritizes sample efficiency, i.e. in converging to a high-quality solution with a few number of iterations. Although scalarized single-objective approaches (e.g., a weighted sum of constrained expected improvements) could produce similar acquisition functions, they would essentially reduce to a simplified version of our framework. In contrast, cEHVI follows a fundamentally different optimization strategy aimed at broadly covering the Pareto front. This comparison highlights the distinction between traditional exploratory multi-objective methods, as exemplified by cEHVI, and the solution-focused trade-off optimization implemented in this work.

\subsubsection{Single objective constrained Bayesian Optimization}
A popular acquisition function for constrained Bayesian optimization with a single objective was proposed by \cite{gardner2014bayesian}, and is called \emph{expected constrained improvement} (EI$_c$). For EI$_c$ it is assumed that each constraint is independent and is also approximated by Gaussian processes. The EI$_c$ is the \emph{expected improvement} (EI) of the approximation of the objective function $\hat{f}(\textbf{x})$ relative to the best feasible objective value $f^*$ observed so far, weighted by the probability that $\textbf{x}$ is a feasible design (\citealp{gardner2014bayesian}). A feasible design is a variable combination $\textbf{x}$ that satisfies $g_i(\mathbf{x}) \leq \lambda_i$, $\forall i \in \{1, \dots, n_c\}$. The probability that the predictions of each constraint $\hat{g}_i(\textbf{x})$ is lower than the limits $\lambda_i$, is known as the probability of feasibility (PF) for design $\textbf{x}$.

\begin{equation} \label{Eq. full cEI}
\text{EI}_\text{c}(\textbf{x}) = \text{EI}(\textbf{x})\text{PF}(\textbf{x})
\end{equation}

The equations for calculating the components of $\text{EI}_c(\textbf{x})$ is given in Eq. \eqref{Eq:cEI}.

\begin{subequations}\label{Eq:cEI}
\begin{align}  
    \text{EI}(\textbf{x}) = & (f^* - \hat{\mu}_f(\textbf{x}))\Phi\left(\frac{f^* - \hat{\mu}_f(\textbf{x})}{\hat{\sigma}_f(\textbf{x})}\right)+\hat{\sigma}_f(\textbf{x})\phi\left(\frac{f^* - \hat{\mu}_f(\textbf{x})}{\hat{\sigma}_f(\textbf{x})}\right) \label{Eq.EI} \\
    \text{PF}(\textbf{x}) = & \prod_{i = 1}^{n_c}\Phi\left(\frac{\lambda_i - \hat{\mu}_{gi}(\textbf{x})}{\hat{\sigma}_{gi}(\textbf{x})}\right) \label{Eq.PF}
\end{align}
\end{subequations}

Here, it is assumed that each of the constraints $\hat{g}_i(x)$ are independent GPs, with mean $\hat{\mu}_{gi}(x)$ and variance $\hat{\sigma}_{gi}^2$. Variable $f^*$ is the best observed objective value so far that satisfies the constraints, $\hat{\mu}_f(\textbf{x})$ and ${\hat{\sigma}_f}^2(\textbf{x})$ are, respectively, the predictive mean and variance of the objective function approximation $\hat{f}(\textbf{x})$. Function $\Phi$ is the cumulative density function (CDF) of a standardized normal distribution, and function $\phi$ is the probability density function (PDF) of a standardized normal distribution.

\subsubsection{Multi-objective constrained Bayesian Optimization}
\subsubsection*{Adopted approach (PTMOO)}
\begin{figure}[H]
    \centering
    \includegraphics[width=0.40\textwidth]{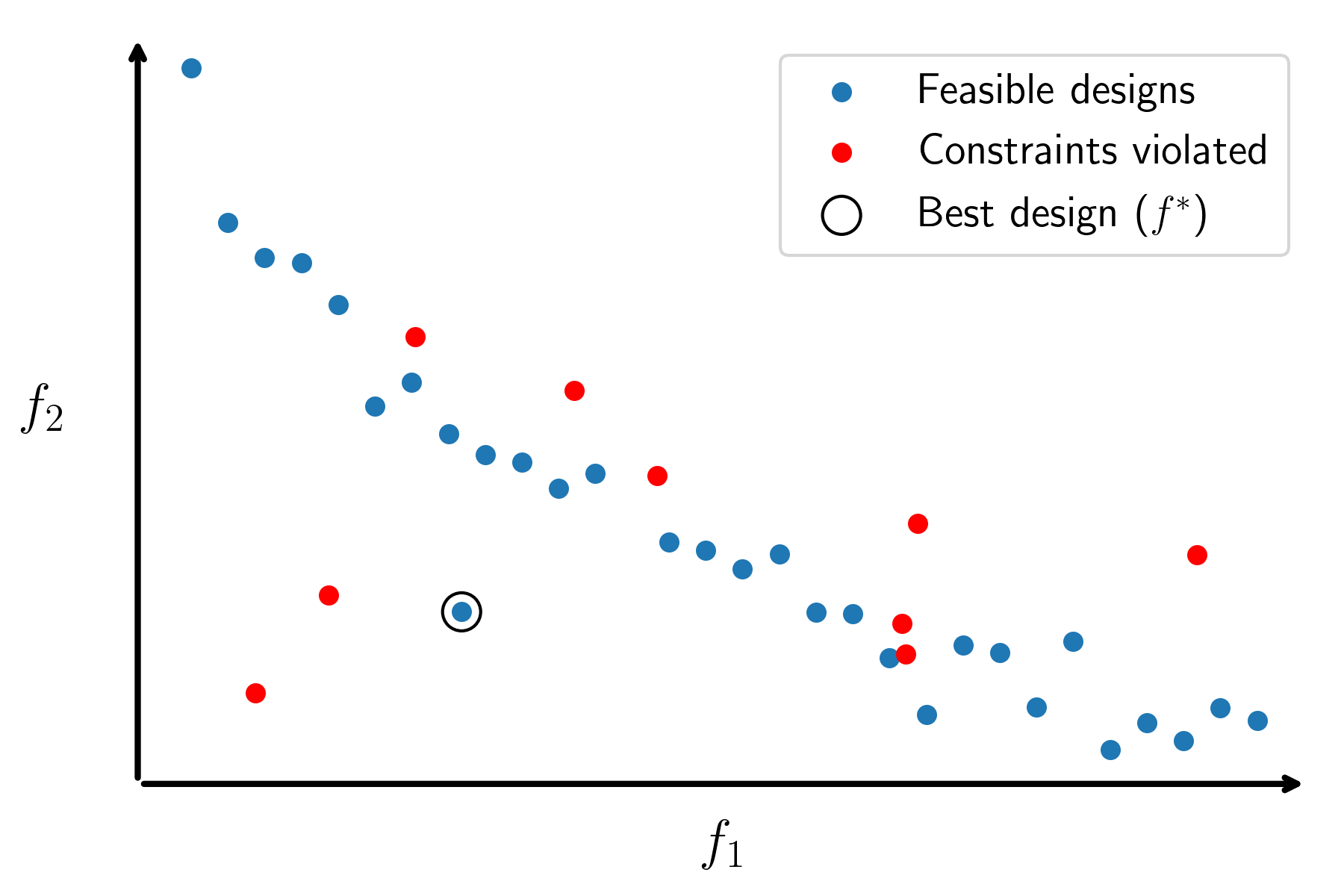}
    \caption{Illustration of best observed design for process with 2 objective functions}
    \label{fig:Optimum_design}
\end{figure}

Within our constrained multi‐objective optimization framework, we compute an acquisition function $\text{EI}_c^{(k)}$ for each of the $n_f$ objective functions. The constrained expected improvement is evaluated relative to a specified \emph{preferred tradeoff}, which we define as the feasible design that yields the most desirable balance among objectives based on a normalized aggregate measure. We use the term \emph{preferred} to emphasize that the optimization process is directed toward a specific solution in the objective space, rather than aiming to reconstruct the full Pareto front. Importantly, this preference is not explicitly defined by the user, e.g. via weights, but arises from a default strategy that assigns equal importance to all objectives. Specifically, we seek a preferred combination of objective function values $\boldsymbol{f}^*(\textbf{x}^*_N) = \left[f^*_1(\textbf{x}^*_N),f^*_2(\textbf{x}^*_N),...,f^*_{n_f}(\textbf{x}^*_N) \right]$. Where $\textbf{x}^*_N \in \textbf{X}$ denotes the \emph{preferred design} at iteration $N$. This is defined as the feasible combination of design variables that minimizes the sum of normalized objective function values:
\begin{subequations} \label{Eq. XN_opt}
\begin{align} 
\textbf{x}_N^* = \underset{\textbf{x} \in \textbf{X}}{\text{arg min}} &\; \sum_{k=1}^{n_f} \frac{f_k(\textbf{x})-f_k^{\text{min}}}{f_k^{\text{max}}-f_k^{\text{min}}} \\
\mathrm{s.t}\quad& g_i(\textbf{x}) \leq \lambda_i, \quad i = 1, 2,...,n_c
\end{align}
\end{subequations}
Here, $f_k^{\min}$ and $f_k^{\max}$ are the minimum and maximum observed values of objective function $f_k$, evaluated across all sampled designs, both feasible and infeasible. Normalization ensures that all objectives are treated with equal importance when computing the aggregate. In cases where differing priorities are desired, weighted normalization can be applied to reflect the relative importance of each objective. Figure~\ref{fig:Optimum_design} illustrates the selection of the best, or preferred, feasible design in a two-objective scenario with equal weighting, as used in Eq. \eqref{Eq. XN_opt}.

\begin{figure}[H]
    \centering
    \includegraphics[width=0.40\textwidth]{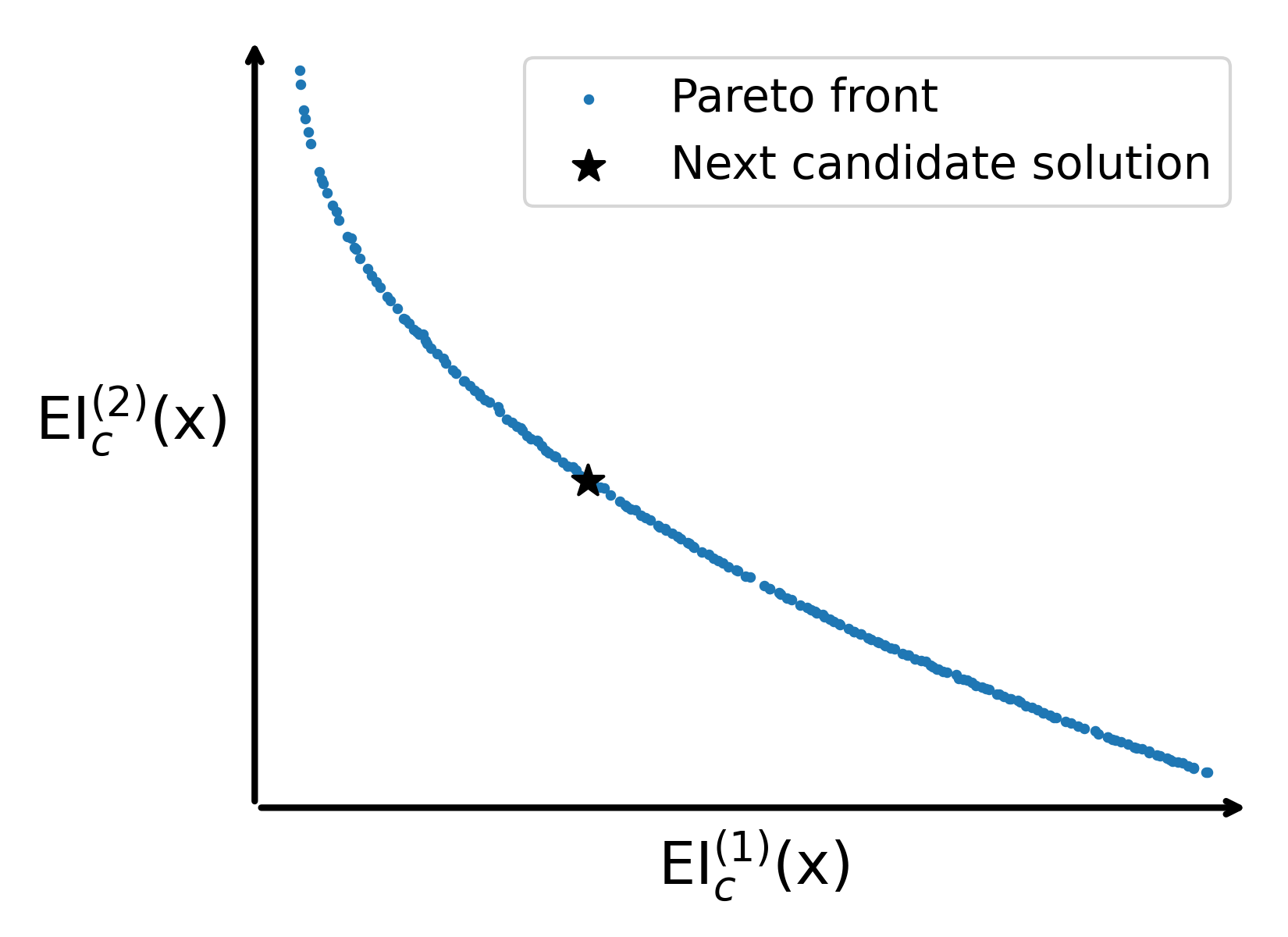}
    \caption{Pareto front constructed using two acquisition functions and the PTMOO candidate solution on the Pareto front using  Eq. \eqref{Eq.OptimumParetoPoint}}
    \label{fig:ParetoFront}
\end{figure}

A Pareto front approximation is then calculated using the multi-objective genetic algorithm \emph{NSGA-II} implemented in the Python package \emph{pymoo} (\citealp{pymoo}). See the article by \cite{deb2002fast} for further information on the NSGA-II algorithm. The objectives of the applied NSGA-II are to maximize the constrained acquisition function EI$_c^{(k)}$ for each of the $n_f$ number of objectives. An illustration of a Pareto front for two acquisition functions is shown with blue dots in Figure \ref{fig:ParetoFront}. When the Pareto front has been determined, the next candidate design, $\textbf{x}_{N+1}$, is determined as the variable combination $\textbf{x}$ with the maximum sum of squared expected relative improvement  of the variable combinations in the the Pareto front $\textbf{X}_P$, which is calculated as:
\begin{equation} \label{Eq.OptimumParetoPoint}
\textbf{x}_{N+1} = \underset{\textbf{x} \in \textbf{X}_P}{\text{arg max}} \; \sum_{k=1}^{n_f}\left(\frac{\text{EI}_c^{(k)}(\textbf{x})}{f^*_k(\textbf{x}_N^*)} \right)^2 \\
\end{equation}

The next candidate design, $\textbf{x}_{N+1}$, is chosen to maximize the aggregate relative improvement over the current preferred solution. It is important to note that $\textbf{x}_{N+1}$ denotes the next candidate design to be evaluated, not necessarily the new preferred design. The actual preferred design at iteration $N+1$, denoted $\textbf{x}^*_{N+1}$, is determined only after evaluating the true objectives and constraints of $\textbf{x}_{N+1}$. We will refer to this selection rule as PTMOO for the remainder of the paper. By using relative improvement, each objective is treated with equal weight while driving the largest possible advancement from the existing preferred design. In this article a single point is selected. Thus, one could also skip calculating the Pareto front approximation and select the next candidate design by simply maximizing Eq. \eqref{Eq.OptimumParetoPoint} for the entire possible design set $\textbf{X}\in \mathbb{R}^d$. The two-stage approach offers three key advantages over "blind" single-objective maximization:
\begin{enumerate}
    \item $\textbf{Decomposes a hard global search into tractable pieces}$\\
        $\text{EI}_c^{(k)}(\textbf{x})$ may be highly multi-modal, so direct global optimization may get stuck in poor local maxima. By first using NSGA-II to approximate the Pareto front of the problem we efficiently identify a diverse set of high-quality candidate points.
    \item $\textbf{Multiple candidate solutions}$\\
        Even though we ultimately choose one $\textbf{x}$ as the next candidate design in this article, the Pareto front approximation can be used to determine multiple candidate points for which to perform new analyzes.
    \item $\textbf{Insight into trade‐offs at each iteration}$\\
        The approximate Pareto front reveals how improvements in one objective affect others, offering valuable insights into trade-offs and guiding early stopping decisions.
\end{enumerate}

Figure \ref{fig:ParetoFront} illustrates the Pareto front approximation of a process with two objective functions, and the location of the next candidate design determined using Eq. \eqref{Eq.OptimumParetoPoint}.

Algorithm \ref{Alg. Efficient_cMOBO} summarizes the proposed efficient multi-objective constrained Bayesian optimization procedure, which incorporates a user-specified preferred trade-off for the objective functions. In our implementation, the full set of hyperparameters is trained in each iteration when approximating the constraints and objectives. We do not define a hard threshold on the preference of KPLS over GPR without hyperparameter reduction. \cite{bouhlel2016improving} showed that when more than ten design variables were considered, KPLS generally matched or outperformed standard GPR. Similarly, we do not prescribe a strict cutoff for applying POD based on output dimensionality, yet we recommend its use when seeking to approximate the full solution set from FEM or CFD analyses.

\begin{algorithm}[H]
\caption{Efficient Multi-Objective Constrained Bayesian Optimization Process}
\begin{algorithmic}[1]
\State \textbf{Input:} Bounds of design variables $[\textbf{X}_{min}, \textbf{X}_{max}]$, initial sampling budget $N_s$, constraint limits $\boldsymbol{\lambda}$
\State \textbf{Output:} Preferred design $\textbf{x}^*$
\State Generate initial sample variables $\textbf{X}_0$ using Latin Hypercube Sampling (LHS) with budget size $N_s$
\State Evaluate initial observations:
$\mathcal{D}_0 = \{ (\textbf{x}_j, \boldsymbol{f}(\textbf{x}_j), \boldsymbol{g}(\textbf{x}_j) )\}_{j=1}^{N_s} = (\textbf{X}_0, \textbf{Y}_{\mathcal{D}_0})$
\State Determine initial preferred design  $\textbf{x}_0^*$ using Eq. \eqref{Eq. XN_opt}
\State Set \( \text{converged} \gets \textbf{False} \)
\State Set iteration counter \( N \gets 0 \)

\While{converged = $\textbf{False}$}
    \For{each objective $k = 1$ to $n_f$}
        \If{Output dimension of $\boldsymbol{f}_k(\textbf{X}_N)$ is large}
            \State Apply Proper Orthogonal Decomposition (POD) to $\boldsymbol{f}_k(\textbf{X}_N)$ for dimension reduction
        \EndIf
        \If{Number of design variables is high}
            \State Approximate $\boldsymbol{f}_k(\textbf{X}_N)$ using KPLS
        \Else
            \State Approximate $\boldsymbol{f}_k(\textbf{X}_N)$ using GPR
        \EndIf
    \EndFor
    \For{each constraint $i = 1$ to $n_c$}
        \If{Output dimension of $\boldsymbol{g}_i(\textbf{X}_N)$ is large}
            \State Apply Proper Orthogonal Decomposition (POD) to $\boldsymbol{g}_i(\textbf{X}_N)$ for dimension reduction
        \EndIf
        \If{Number of design variables is high}
            \State Approximate $\boldsymbol{g}_i(\textbf{X}_N)$ using KPLS
        \Else
            \State Approximate $\boldsymbol{g}_i(\textbf{X}_N)$ using GPR
        \EndIf
    \EndFor
     \State Compute Pareto front using Eq. \eqref{Eq. full cEI} as acquisition functions for the \( n_f \) objectives
    \State Select next candidate design $\textbf{x}_{N+1}$ from Pareto front using Eq. \eqref{Eq.OptimumParetoPoint}
    \State Evaluate objectives and constraints:  
          $\{{f}_1(\textbf{x}_{N+1}),\ldots,{f}_{n_f}(\textbf{x}_{N+1}), {g}_1(\textbf{x}_{N+1}),\ldots,{g}_{n_c}(\textbf{x}_{N+1}) \}$
    \State Update observations: $\mathcal{D}_{N+1} \gets \mathcal{D}_N \cup (\textbf{x}_{N+1}, \boldsymbol{f}(\textbf{x}_{N+1}), \boldsymbol{g}(\textbf{x}_{N+1}) )$
    \State Determine preferred design $\textbf{x}_{N+1}^*$ using Eq. \eqref{Eq. XN_opt}
    \If{Convergence criteria met}
        \State \( \text{converged} \gets \textbf{True} \)
    \EndIf
    \State $N \gets N + 1$
\EndWhile

\State \textbf{Return} $\textbf{x}_{N+1}^*$

\end{algorithmic}
\label{Alg. Efficient_cMOBO}
\end{algorithm}

\subsubsection*{Constrained Expected Hypervolume Improvement (cEHVI)}
To compare the results obtained using Algorithm \ref{Alg. Efficient_cMOBO}, results obtained using the \emph{constrained Expected Hypervolume Improvement} (cEHVI) acquisition function is applied. The cEHVI is an extension of the cEI defined in Eq. \eqref{Eq:cEI} (\citealp{daulton2020differentiable}). Given a finite approximation set of $n_\mathcal{P}$ points to a Pareto front $\mathcal{P} = \left\{ \boldsymbol{f}^{(1)},\ldots, \boldsymbol{f}^{(n_{\mathcal{P}})}\right\} \subset\mathbb{R}^{n_f}$, the \emph{hypervolume} (HV) indicator measures the size of the dominated subspace bounded above by a reference point $\boldsymbol{r}\in \mathbb{R}^{n_f}$:
\begin{equation}\label{Eq.HV}
    \text{HV}(\mathcal{P,\boldsymbol{r}})=\lambda_{n_f}\left(\cup_{\boldsymbol{f}\in\mathcal{P}}\left[\boldsymbol{f},\boldsymbol{r}\right]\right),
\end{equation}
where $\lambda_{n_f}$ is the Lebesgue measure. For a vector $\boldsymbol{f}(\textbf{x})\in\mathbb{R}^{n_f}$ the \emph{hypervolume improvement} (HVI) is calculated as:
\begin{equation}\label{Eq.HVI}
    \text{HVI}(\textbf{x},\mathcal{P},\boldsymbol{r}) = \text{HV}(\mathcal{P}\cup\left\{\boldsymbol{f}(\textbf{x}) \right\},\boldsymbol{r})-\text{HV}(\mathcal{P,\boldsymbol{r}})
\end{equation}
When the objectives $\boldsymbol{f}(\textbf{x})$ are approximated by a predictive distribution, e.g. GPR, then for a given set of predictive distributions $p(\boldsymbol{f}|\textbf{x}) = \mathcal{N}\left(\boldsymbol{\mu}(\textbf{x}), \boldsymbol{\sigma}^2(\textbf{x})\right)$, and the Pareto front approximation $\mathcal{P}$, the \emph{expected hypervolume improvement} (EHVI) is defined as:
\begin{equation} \label{Eq.EHVI}
    \text{EHVI}(\textbf{x},\mathcal{P},\boldsymbol{r}) = \int_{\mathbb{R}^{n_f}}
    \text{HVI}(\textbf{x}, \mathcal{P},\boldsymbol{r})\cdot p(\boldsymbol{f}|\textbf{x}) d\boldsymbol{f}(\textbf{x})
\end{equation}
The cEHVI may then be calculated by combining the PF from Eq. \eqref{Eq.PF} with that of EHVI in Eq. \eqref{Eq.EHVI}:
\begin{equation}
    \text{cEHVI}(\textbf{x},\mathcal{P},\boldsymbol{r}) = \text{PF}(\textbf{x})\cdot\text{EHVI}(\textbf{x},\mathcal{P},\boldsymbol{r})
\end{equation}
The next candidate design to be sampled, $\textbf{x}_{N+1}$ is then calculated by maximizing the cEHVI:
\begin{equation} \label{Eq. cEHVI_xn1}
    \textbf{x}_{N+1} = \underset{\textbf{x} \in \textbf{X}}{\text{arg max }} \text{cEHVI}(\textbf{x},\mathcal{P},\boldsymbol{r})
\end{equation}
The cEHVI and the PTMOO frameworks share a common goal, guiding the optimization process toward designs with the most promising improvement. However, they differ fundamentally in how they define and pursue improvement.
\begin{itemize}
    \item PTMOO focuses on a single preferred design. The acquisition function is computed as the expected improvement, conditioned on feasibility, relative to a specific target, the preferred trade-off point $\boldsymbol{f}(\textbf{x}^*_N)$. This leads to a point-based improvement strategy, more exploitative and aligned with user preferences.
    \item cEHVI, on the other hand, computes the expected increase in hypervolume over the entire current approximation of the Pareto front, conditioned on feasibility. This makes it inherently set-based and more diversity-oriented, as it rewards candidate solutions that extend or fill gaps in the front.
\end{itemize}
The objective function values of cEHVI are typically normalized to $[0,1]^{n_f}$ and a reference point $\boldsymbol{r} = \boldsymbol{1}^{n_f}$ used (\citealp{mathern2021multi}). This approach is also followed in this article. Alternatively a reference point using the objective values determined using Eq. \eqref{Eq. XN_opt} could be applied, which would limit the search space and avoid the extreme non-dominated points. As the cEHVI in this article is intended only to compare against the approach outlined in Algorithm \ref{Alg. Efficient_cMOBO}, the typical reference point of $\boldsymbol{r} = \boldsymbol{1}^{n_f}$ is used.

\section{Case-study} \label{Ch. Case-study}
\begin{figure}[H]
    \centering
    \begin{subfigure}{0.3\textwidth}
        \centering
        \includegraphics[width=\linewidth]{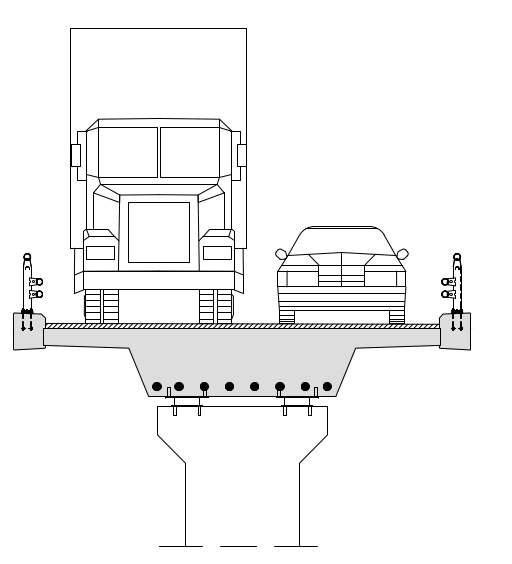}
        \caption{Girder section}
        \label{fig:Bridge_SectionIllustration}
    \end{subfigure}
    \begin{subfigure}{0.65\textwidth}
        \centering
        \includegraphics[width=\linewidth]{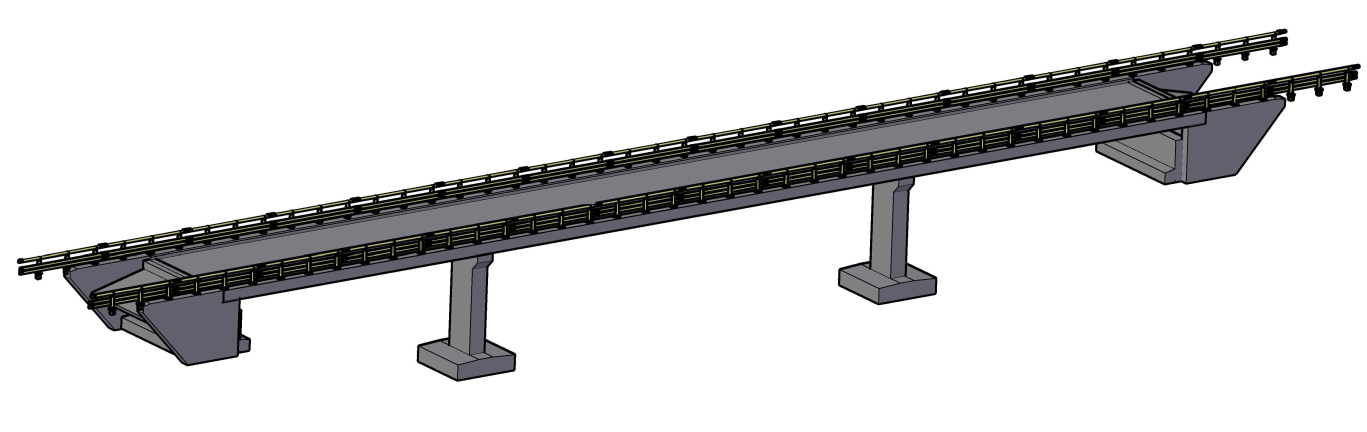}
        \caption{Perspective view of the bridge}
        \label{fig:Bridge_perspectiveView}
    \end{subfigure}
    
    \caption{Illustration of a 3 span post-tensioned concrete road bridge with 2 lanes}
    \label{fig:BridgeIllustration}
\end{figure}

The bridge considered is a three-span, post-tensioned concrete road bridge with 2 traffic lanes. The side spans are assumed to be 20~m each and the central span is 30~m, thus the entire bridge is 70~m. A perspective view and a view of the girder section are illustrated in Figure~\ref{fig:BridgeIllustration}. In this article only the bridge girder is considered during the optimization. 

\subsection{Variables and geometry}
For the case study, 15 design variables are considered, as indicated in Table~\ref{table:Variables}, where the upper and lower bounds of the variables is also provided. These design variables are considered to be the most important for the girder design.
\begin{itemize}[itemsep=1pt, topsep=0pt]
    \item Tendons provide a compressive axial force to the girder and also a moment which counteracts the girder self-weight and additional live loads.
    \item Concrete has a relatively low tensile strength, and the reinforcement is used to strengthen the areas of the girder which is subjected to tensile loads, in addition to reducing crack widths.
    \item The distribution of the reinforcement is partitioned into what is considered as reasonable distribution areas, depending on what part of the section is most subjected to tension or compression.
    \item The girder's height is the most important parameter for the girder's bending stiffness about the weak axis.
\end{itemize}

\begin{table}[h!]
    \centering
    \small
    \caption{Variables for the MOO and associated upper and lower bounds}
    \label{table:Variables}
    \begin{tabular}{ l l r r}
    \textbf{Variable} & &\multicolumn{1}{c}{\textbf{Lower bound}} & \multicolumn{1}{c}{\textbf{Upper bound}} \\
    \hline
    Tendon strand section area &   &   &\\
    \hspace{2em} Main span & ($A_{TsM}$) & 70.0 $\text{mm}^2$ & 180.0 $\text{mm}^2$\\
    \hspace{2em} Side spans & ($A_{TsS}$)& 70.0 $\text{mm}^2$ & 180.0 $\text{mm}^2$\\
    Tendon inflection point - Side spans$^*$ & ($\alpha_{L,TsS}$) & 0.50 & 0.80\\
    Girder cross section height & ($H$) & 0.80 $\text{m}$ & 2.00 $\text{m}$\\
    Reinforcement area &   &   &   \\
    \hspace{2em} Upper section - Part 1 & ($A_{T1}$) & 3 000 $\text{mm}^2$ & 10 000 $\text{mm}^2$\\
    \hspace{2em} Upper section - Part 2 & ($A_{T2}$) & 3 000 $\text{mm}^2$ & 10 000 $\text{mm}^2$\\
    \hspace{2em} Upper section - Part 3 & ($A_{T3}$) & 3 000 $\text{mm}^2$ & 10 000 $\text{mm}^2$\\
    \hspace{2em} Lower section - Part 1 & ($A_{B1}$) & 3 000 $\text{mm}^2$ & 10 000 $\text{mm}^2$\\
    \hspace{2em} Lower section - Part 2 & ($A_{B2}$) & 3 000 $\text{mm}^2$ & 10 000 $\text{mm}^2$\\
    \hspace{2em} Lower section - Part 3 & ($A_{B3}$) & 3 000 $\text{mm}^2$ & 10 000 $\text{mm}^2$\\
    \hspace{2em} Lower section - Part 4 & ($A_{B4}$) & 3 000 $\text{mm}^2$ & 10 000 $\text{mm}^2$\\
    Reinforcement distribution length$^*$ &   &    &  \\
    \hspace{2em} Upper section - Part 1 & ($\alpha_{L,T1}$) & 0.10 & 0.95\\
    \hspace{2em} Upper section - Part 3 & ($\alpha_{L,T3}$) & 0.10 & 0.95\\
    \hspace{2em} Lower section - Part 2 & ($\alpha_{L,B2}$) & 0.10 & 0.95\\
    \hspace{2em} Lower section - Part 4 & ($\alpha_{L,B4}$) & 0.10 & 0.95\\
    \hline
    \end{tabular}
    % \vspace{0.05cm} % Space before notes
    {\footnotesize * The lengths are relative to the respective span lengths.}
\end{table}

\begin{figure}[H]
    \centering
    \begin{subfigure}{0.7\textwidth}
        \centering
        \includegraphics[width=\linewidth]{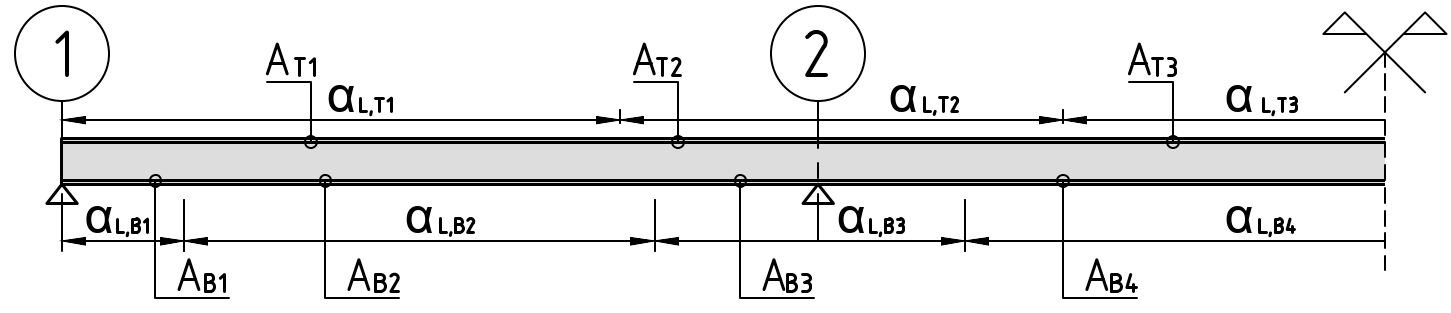}
        \caption{Reinforcement variables along the girder}
        \label{fig:ReinforcementAlongBridge}
    \end{subfigure}
    \begin{subfigure}{0.7\textwidth}
        \centering
        \includegraphics[width=\linewidth]{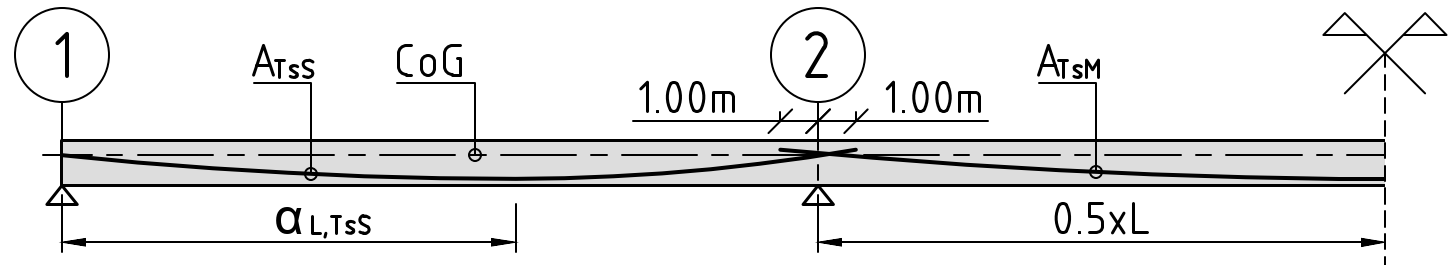}
        \caption{Tendon variables along the girder}
        \label{fig:TendonsAlongBridge}
    \end{subfigure}
    
    \caption{Variables along the girder}
    \label{fig:VariablesAlongBridge}
\end{figure}

\begin{figure}[H]
    \centering
    \begin{subfigure}{0.40\textwidth}
        \centering
        \includegraphics[width=\linewidth]{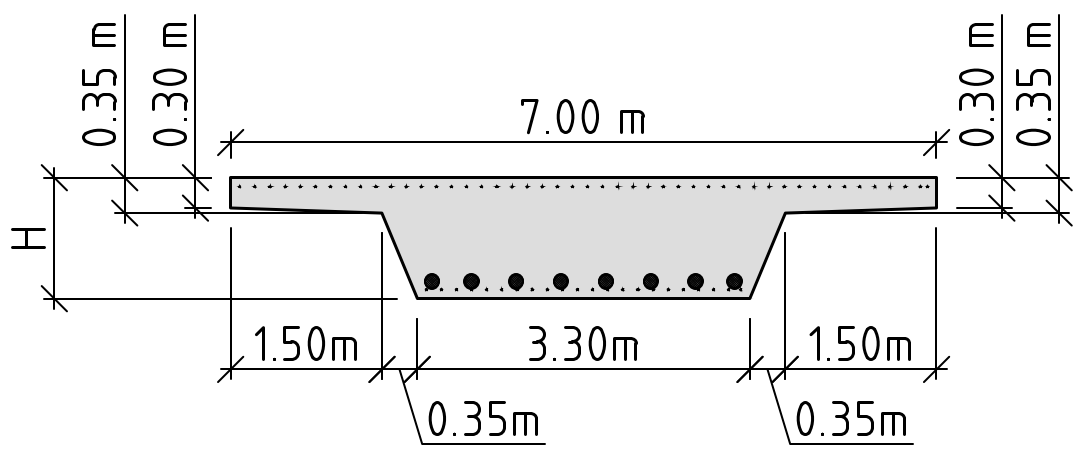}
        \caption{Geometry}
        \label{fig:GirderSectionGeometry}
    \end{subfigure}
    \begin{subfigure}{0.35\textwidth}
        \centering
        \includegraphics[width=\linewidth]{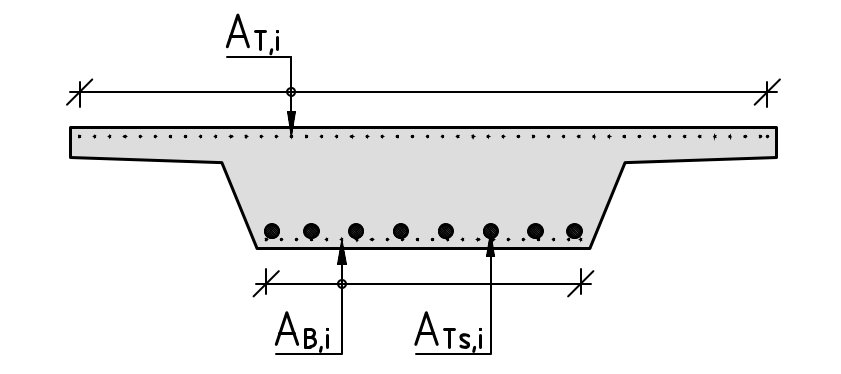}
        \caption{Reinforcement and tendons}
        \label{fig:GirderTendonsReinforcement}
    \end{subfigure}
    
    \caption{Girder cross-sectional variables and geometry}
    \label{fig:GirderVariablesGeometry}
\end{figure}

The position of the variables along the bridge girder is illustrated in Figure~\ref{fig:VariablesAlongBridge}, and the cross-sectional geometry and the position of the variables in the girder's cross-section is illustrated in Figure~\ref{fig:GirderVariablesGeometry}.

In Figure \ref{fig:GirderSectionGeometry} the geometric values of the girder are shown. As indicated in the figure, only the height of the cross-section is considered as a variable. The upper width must be constant to accommodate a certain number of traffic lanes, and the bottom width must be constant to ensure adequate spacing for the tendons. Remaining geometry is kept fixed as lateral force on the girder is not considered, and the applied values ensures a realistic design.

\subsubsection{Materials}
The material quality assumed in the girder of the concrete, reinforcement bars and post-tension system is rendered in Table \ref{table:GirderMaterial}. 

\begin{table}[h!]
    \centering
    \caption{Materials applied in the bridge girder}
    \label{table:GirderMaterial}
    \begin{tabular}{ l r}   
    \textbf{Material} & \textbf{Quality}\\
    \hline
    Concrete &  B45 \\
    Reinforcement bars & B500NC \\ 
    Post-tension tendons & Y1860S7 \\
    \hline
    \end{tabular}
\end{table}

\subsubsection{Loading}
The girder is subjected to permanent and variable loads. The permanent loads are comprised of self-weight, creep, shrinkage, super dead-loads and the post-tension force. The variable loads comprise of traffic and temperature loads. The total carriage width of the girder is 7.0 m, which enables two notional lanes. The edge beams, on which the railing is mounted, are assumed suspended to the girders cantilevers. Table \ref{table:AppliedLoads} shows the applied loads. As the cross section and amount of post-tension varies, references to the applied regulations are given when relevant.

\begin{table}[h!]
    \centering
    \caption{Applied loads}
    \label{table:AppliedLoads}
    \begin{tabular}{ l r @{\hskip 3pt} l}   
    \multicolumn{3}{l}{\textbf{Permanent loads}} \\
    \hline
    \quad Selfweight concrete &  25.00 & kN/m$^3$ \\
    \quad Surfacing &  2.50 & kN/m$^2$ \\
    \quad Edge beams &  8.75 & kN/m \\
    \quad Railing & 0.50 & kN/m \\
    \quad Creep & \multicolumn{2}{l}{Annex B.1 in \cite{EC2-1-1}} \\
    \quad Shrinkage & \multicolumn{2}{l}{Annex B.2 in \cite{EC2-1-1}} \\
    \quad Post-tension & \multicolumn{2}{l}{Section 5.10 in \cite{EC2-1-1}} \\
    \multicolumn{3}{l}{\textbf{Variable loads}}\\
    \hline
    \quad Uniform traffic - Lane 1 & 5.40 & kN/m$^2$ \\
    \quad Uniform traffic - Lane 2 & 2.50 & kN/m$^2$ \\
    \quad Uniform traffic - Remaining area & 2.50 & kN/m$^2$ \\
    \quad Tandem traffic - Lane 1 & 2$\times$300 & kN \\
    \quad Tandem traffic - Lane 2 & 2$\times$200 & kN \\
    \quad Max. air temperature &  36.0 & \textdegree{}C \\ 
    \quad Min. air temperature & -30.0 & \textdegree{}C \\
    \hline
    \end{tabular}
\end{table}

\subsection{Load combinations}\label{LoadCombs}
The internal loads from the FEM simulations are combined to what is known as \emph{load combinations}, which reflects load combination to which the structure is, with different likelihood and duration, subjected to during it's lifetime. In Table \ref{table:ULS_loadFactors} the load factors for the Ultimate Limit State (ULS) is shown.
\begin{table}[h!]
    \centering
    \small
    \caption{Ultimate Limit State (ULS) load factors}
    \label{table:ULS_loadFactors}
    \begin{tabular}{l | c c c | c c}   
    \textbf{Dominant load}  & \multicolumn{3}{c|}{\textbf{Permanent loads*}} & \multicolumn{2}{c}{\textbf{Variable loads}} \\
     & \textbf{Selfweight} & \textbf{Deformation} & \textbf{Tendons} & \textbf{Traffic} & \textbf{Temperature} \\
    \hline 
    Selfweight  & 1.35/1.00 & 1.00/0.00 & 1.10/0.90 & 0.95 & 0.84 \\
    Traffic     & 1.20/1.00 & 1.00/0.00 & 1.10/0.90 & 1.35 & 0.84 \\
    Temperature & 1.20/1.00 & 1.00/0.00 & 1.10/0.90 & 0.95 & 1.20 \\
    \hline
    \end{tabular}
    {\footnotesize * The load factors are given as unfavorable/favorable.}
\end{table}

For the Serviceability Limit State (SLS), which are load combinations associated to the serviceability of the structure, three main load combinations are considered:
\begin{itemize}[itemsep=1pt, topsep=0pt]
    \item Char: Characteristic
    \item Freq: Frequent
    \item Perm: Quasi-permanent
\end{itemize}
The load factors for the load combinations with associated dominant load are shown in Table \ref{table:SLS_loadFactors}.
\begin{table}[h!]
    \centering
    \small
    \caption{Serviceability Limit State (SLS) load factors}
    \label{table:SLS_loadFactors}
    \begin{tabular}{l | l | c c c | c c}   
    \textbf{Comb.} & \textbf{Dominant load}  & \multicolumn{3}{c|}{\textbf{Permanent loads}} & \multicolumn{2}{c}{\textbf{Variable loads}} \\
     & & \textbf{Selfweight} & \textbf{Deformation*} & \textbf{Tendons} & \textbf{Traffic} & \textbf{Temperature} \\
    \hline 
    \multirow{2}{*}{Char.} & Traffic     & 1.00 & 1.00/0.00 & 1.00 & 1.00 & 0.70 \\
                           & Temperature & 1.00 & 1.00/0.00 & 1.00 & 0.70 & 1.00 \\
    \hline
    \multirow{2}{*}{Freq.} & Traffic     & 1.00 & 1.00/0.00 & 1.00 & 0.70 & 0.20 \\
                           & Temperature & 1.00 & 1.00/0.00 & 1.00 & 0.20 & 0.60 \\
    \hline
    \multirow{2}{*}{Perm.} & Traffic     & 1.00 & 1.00/0.00 & 1.00 & 0.50 & 0.20 \\
                           & Temperature & 1.00 & 1.00/0.00 & 1.00 & 0.20 & 0.50 \\
    \hline
    \end{tabular}
    {\footnotesize * The load factors are given as unfavorable/favorable.}
\end{table}

\newpage
\subsection{Constraints} \label{Ch. Constraints}
The constraints considered for the MOO of the bridge girder is given in Table \ref{table:Constraints}. The constraints are associated to both the Ultimate and Serviceability limit state (ULS and SLS) load combinations, in addition to deflections solely due to the traffic loads. For the constraints concerning the reinforcement, it is separated into reinforcement at the upper and lower part of the girder section, thus a total of nine constraints are considered.
\begin{table}[h!]
    \centering
    \small
    \caption{Constraints and their limits considered for the MOO of the bridge girder}
    \label{table:Constraints}
    \begin{tabular}{ l r | l}
    Ultimate Limit State (ULS) & & Reference\\
    \hline
    \hspace{2em} Tendon strain & 10.0 \textperthousand & Section NA.3.3.6 in \cite{EC2-1-1}\\
    \hspace{2em} Concrete compressive strain & -3.5 \textperthousand & Section 3.1.7 in \cite{EC2-1-1}\\
    \hspace{2em} Reinforcement strain & 30.0 \textperthousand & Table NA.3.5(901) in \cite{EC2-1-1}\\
    Serviceability Limit State (SLS) & &\\
    \hline
    \hspace{2em} Concrete compressive stress (SLS-Char.) & -27.0 MPa & Section 7.2(2) in \cite{EC2-1-1}\\
    \hspace{2em} Concrete compression at tendons (SLS-Perm.) & 0.0 MPa & Table NA.7.1N in \cite{EC2-1-1}\\
    \hspace{2em} Crack width (SLS-Freq.) & 0.13 mm & Table NA.7.1N in \cite{EC2-1-1}\\
    Maximum deflection due to traffic$^*$ & $\text{L}_i/350$ & Section 3.5 in \cite{N400}\\
    \hline
    \end{tabular}
    {\footnotesize * $\text{L}_i$ is the span length of the span where the deflection is considered.}
\end{table}

The constraints are derived from numerical analyses and the results from along the entire bridge is used. Thus, the constraints are represented by high-dimensional, functional outputs.

\subsection{Objectives}
The goal of this MOO process is to minimize both the monetary and environmental costs associated to the materials of the bridge girder. Though this case study is limited to two objectives, expanding a MOO process to additional objectives is straightforward; see e.g. \cite{yepes2024hybrid}. The environmental cost is for the case study determined based on $\text{CO}_2$ equivalent emissions derived from the life cycle assessment (LCA) of the materials used in the structure, with data sourced from the Norwegian Public Roads Administration’s LCA tool, \emph{VegLCA} (\citealp{VegLCA}). Similarly, the monetary cost is calculated using material costs considered representative of Norwegian infrastructure projects. The values applied to the different materials are provided in Table \ref{table:ObjectiveCosts}.
 
\begin{table}[h!]
    \centering
    \small
    \caption{Material monetary and environmental costs}
    \label{table:ObjectiveCosts}  
    \begin{tabular}{ l r @{\hskip 2pt} l r @{\hskip 2pt} l}
    \textbf{Material} & \multicolumn{2}{c}{\textbf{Environmental cost}} & \multicolumn{2}{c}{\textbf{Monetary cost}} \\
    \hline
    Concrete (grade B45) & 360 & $\text{kgCO}_2\text{-eq/m}^3$ & 4 200 & $\text{NOK/m}^3$ \\
    Normal reinforcement (grade B500NC) & 4 443 & $\text{kgCO}_2\text{-eq/m}^3$ & 295 945 & $\text{NOK/m}^3$ \\
    Tendons (grade Y1860) & 12 & $\text{kgCO}_2\text{-eq/mMN}$ & 700 & $\text{NOK/mMN}$ \\
    \hline
    \end{tabular}
\end{table}

\section{Optimizing the girder}
\subsection{Numerical Analyses}
The bridge girder is analyzed using a \emph{Finite Element Method} (FEM) model. The FEM model is a beam element model, which is modeled and analyzed using a software called NovaFrame, which is an in-house software at \cite{NovaFrame6_2}. The girder is modeled using 60 beam elements (61 nodes). The analyses are performed using linear-elastic theory to derive sectional forces. The sectional forces from the different permanent and variable loads are then combined for the various limit states, and automatically sorted to determine the most adverse load combinations. The software then has a module that allows the stresses and strains in the concrete, reinforcement, and tendons at the section to be calculated by determining the equilibrium between the external and internal forces of a monotone stress field using a discreet approximation to Green’s theorem.

\subsection{Approximating Simulation Results}
\begin{figure}[H]
    \centering
    \begin{subfigure}{0.45\textwidth}
        \centering
        \includegraphics[width=\linewidth]{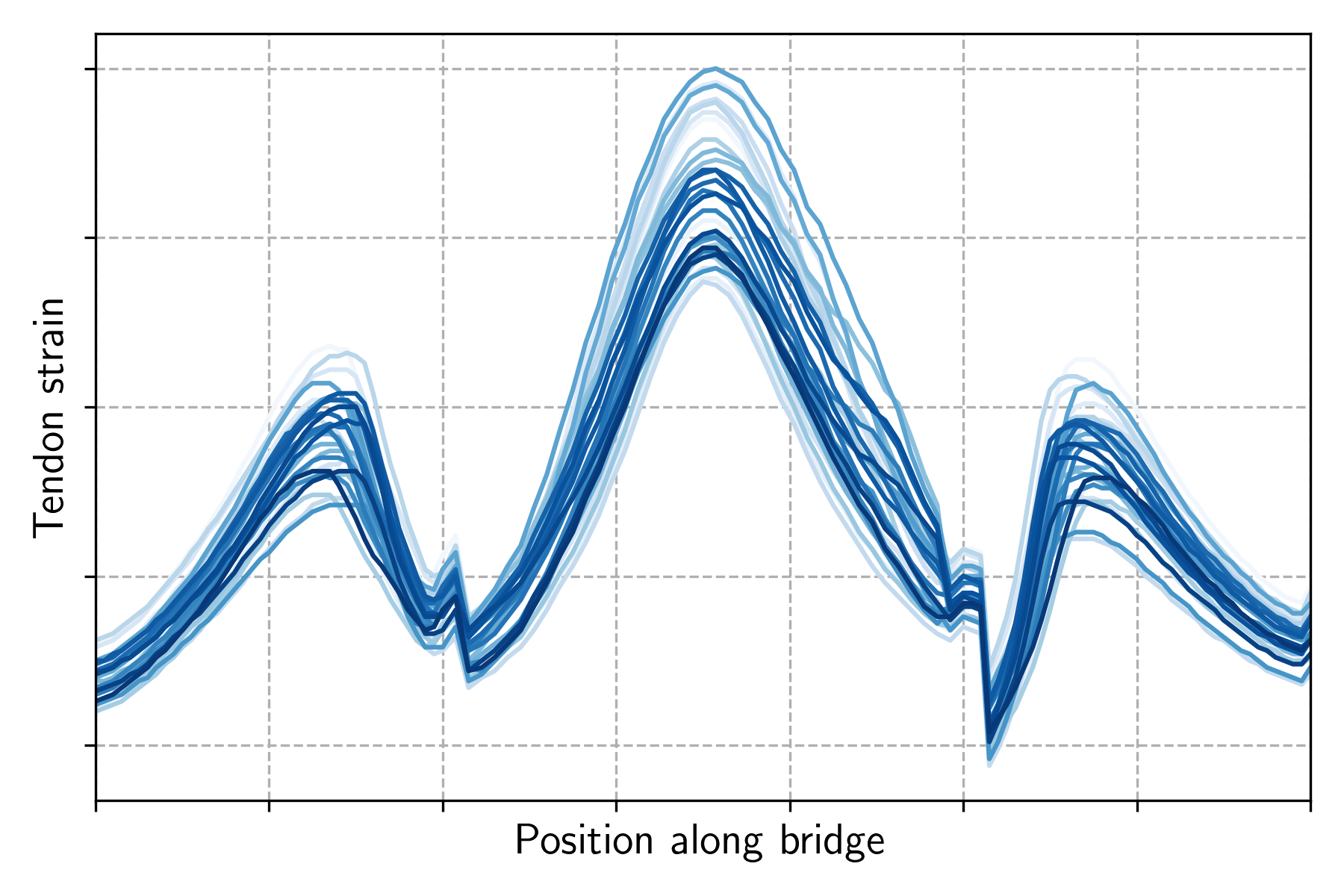}
        \caption{Tendon strain}
        \label{fig:Tendon_response}
    \end{subfigure}
    \begin{subfigure}{0.45\textwidth}
        \centering
        \includegraphics[width=\linewidth]{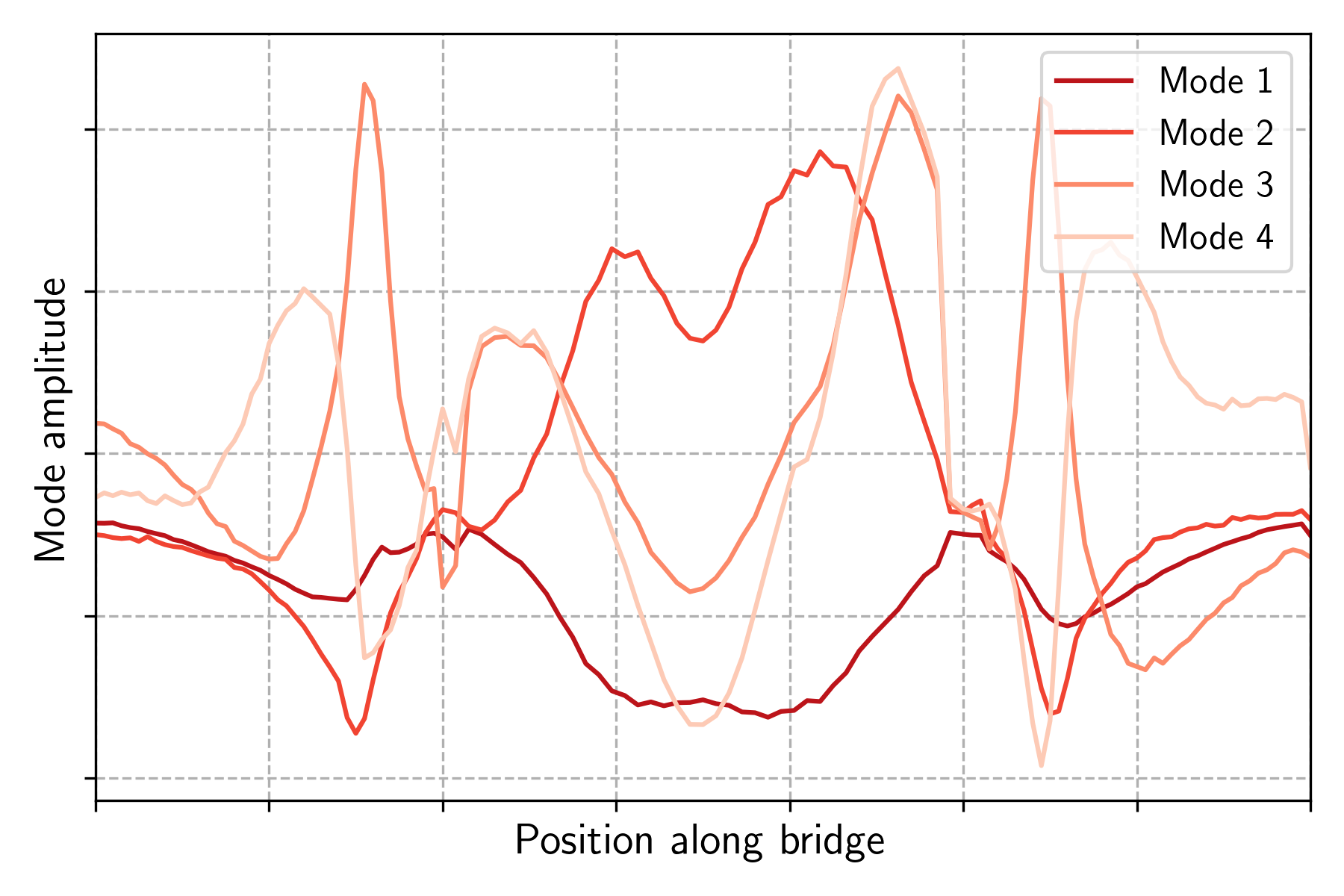}
        \caption{POD mode shapes}
        \label{fig:Tendon_POD_modes}
    \end{subfigure}
    
    \caption{Tendon strain and corresponding POD mode shapes of the 30 initial observations along the bridge. The 4 POD mode shapes correspond to the necessary number of modes with ${\epsilon_{POD}}^2 = 0.01$.}
    \label{fig:Tendon_training_data}
\end{figure}

LHS with normalized distributions for each design variable is carried out using 30 samples. Hence, $N_s = 30$ independent analyses are performed, each corresponding to a unique combination of design variables $\mathbf{x}$, to determine the initial observation set $\mathcal{D}_0$. As described in Section \ref{Ch. Constraints}, nine constraints are considered. For each constraint $i$, the numerical simulations produce results at the 61 nodal points and the 60 element midpoints, yielding $N_h = 61 + 60 = 121$ spatial evaluations. The constraints along the girder are given by:
\begin{equation} \label{eq:utilization_constraints}
    g_i = 
    \begin{cases}
        UR_i = \dfrac{F_{Ed,i}}{F_{Rd,i}}, & i \neq \textit{concrete compression at tendons}\\[0.75em]
        F_{Ed,i},     & i = \textit{concrete compression at tendons}
    \end{cases} \\[0.5em]
\end{equation}
Where $UR_i$ is used as an abbreviation of the \emph{utilization ratio} for constraint $i$. \(F_{Ed,i}\) is the subjected load for constraint \(i\), \(F_{Rd,i}\) its capacity (see Table~\ref{table:Constraints}), and for the constraint “concrete compression at tendons” we simply take the raw analysis result. The constraints are limited as:
\begin{equation}
    g_i \le \lambda_i,\quad
    \lambda_i =
    \begin{cases}
        1.0, & i \neq \textit{concrete compression at tendons}\\
        0.0, & i = \textit{concrete compression at tendons}
    \end{cases}
\end{equation}
 This signifies that the subjected load must be lower than the capacity, and for the constraint "concrete compression at tendons" the concrete must be in compression. The constraints are then approximated using POD with GPR to interpolate for unobserved variable combinations. As detailed in Section \ref{Ch.POD}, the number of POD modes is selected using Eq. \ref{Eq: Number of POD modes} with ${\epsilon_{POD}}^2 = 0.01$. Because each constraint has different POD modes, the required number of POD modes varies across constraints. Figure \ref{fig:Tendon_training_data} illustrates the tendon strain along the bridge for the initial 30 observations, together with the first four POD modes which were sufficient to keep the projection error below 0.01 for the initial observations of the tendon strain. In Figure \ref{fig:Tendon_POD_modes}, the modes are arranged in descending order of significance, with “Mode 1” denoting the most significant mode.

To assess the effectiveness of the KPLS method the constraints are approximated using POD with GPR and no kernel hyper-parameter reduction, and using POD with KPLS where the number of components included in the KPLS ranges from 2 to 5.

The objectives $\boldsymbol{f}$, i.e. the monetary and environmental cost, are obtained for the complete girder as scalar values, and are thus not approximated using POD model reduction, though they are approximated using GPR with and without kernel hyperparameter reduction in the same manner as for the constraints. The objective functions have a relatively simple shape and could also be accurately modeled using a surrogate modeling technique that is simpler than GPR. The benefit of using GPR for the objective functions is, as mentioned before, that it allows the optimization process to leverage uncertainties between observations. GPR is also relatively fast to construct when a limited number of simulations are performed, which is typical in engineering design.

The calculations are implemented in Python using TensorFlow (\citealp{tensorflow2015-whitepaper}). The estimation of hyperparameters is performed using the gradient-based L-BFGS algorithm from Scipy, where the gradients are calculated using algorithmic differentiation in TensorFlow (\citealp{scipy}). 

\subsection{Optimization Routine}
The MOO processes described in Section \ref{Ch.OptProc} are performed for 50 iterations, increasing the sample size $N_s$ with one new observation for each iteration. This is repeated 20 times, each run using the same initial set of 30 samples, to assess its consistency. The hyperparameter setting for the genetic MOO algorithm NSGA-II were;
\begin{itemize}[itemsep=1pt, topsep=0pt]
    \item Population size: 200
    \item Number of generations: 100
\end{itemize}
To also assess the effect of the KPLS hyper-parameter reduction, the following GPR approximations are considered:
\begin{itemize}[itemsep=1pt, topsep=0pt]
    \item \textbf{GPR}: No hyper-parameter reduction
    \item \textbf{KPLS}$_h$: KPLS with $\forall h \in \{2,3,4,5\}$ components
\end{itemize}
To implement cEHVI, we employed box-decomposition (\citealp{yang2019efficient}). Optimization of Eq. \eqref{Eq. cEHVI_xn1} was carried out with the genetic algorithm (GA) provided by Pymoo (\citealp{pymoo}), ensuring both MOO routines shared the same solution strategy. The hyperparameters were configured using the same hyperparameters as for the NSGA-II. It was also considered optimizing Eq. \eqref{Eq. cEHVI_xn1} directly using a gradient based optimizer (GBO). This was performed using the L-BFGS GBO algorithm in Scipy (\citealp{scipy}). The GBO employed a multi-start strategy, where 1000 initial designs were generated at each iteration using uniform Latin hypercube sampling (LHS), and the best candidate was selected as the starting point for the optimization.

For the cEHVI and GBO, only GPR with no hyperparameter reduction and KPLS with 5 components were considered.

\subsection{Benchmark}
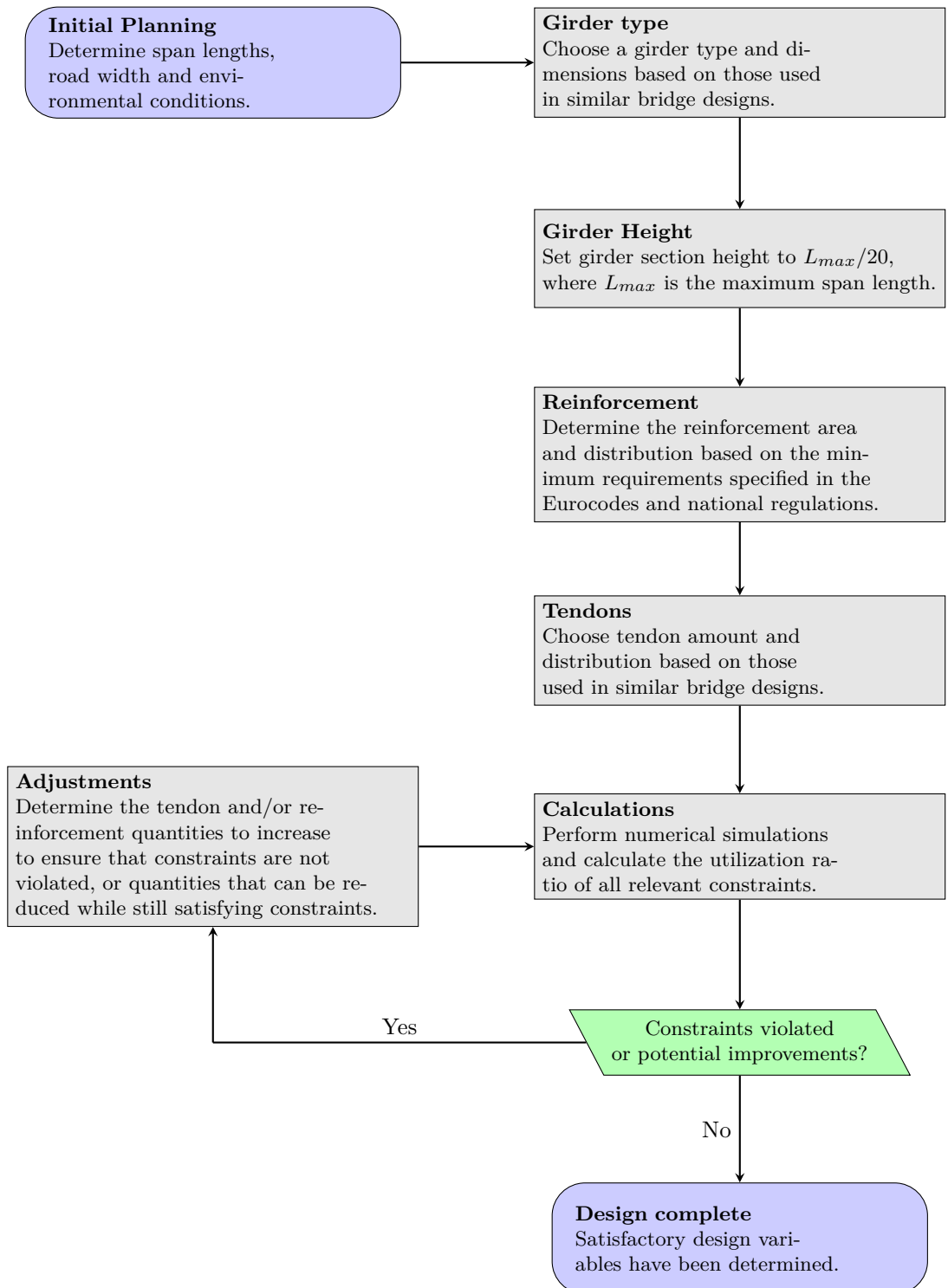
\begin{figure}[H]
    \centering
        \begin{tikzpicture}[node distance=3.0cm]
        
        \node (Initial) [start] {\textbf{Initial Planning}\\ 
            Determine span lengths, road width and environmental conditions.};
        \node (GirderType) [process, right of = Initial, xshift = 5.0cm] {\textbf{Girder type} \\
        Choose a girder type and dimensions based on those used in similar bridge designs.};
        \node (GirderHeight) [process, below of = GirderType] {\textbf{Girder Height} \\
        Set girder section height to $L_{max}/20$, where $L_{max}$ is the maximum span length.};
        \node (Reinforcement) [process, below of = GirderHeight] {\textbf{Reinforcement} \\
        Determine the reinforcement area and distribution based on the minimum requirements specified in the Eurocodes and national regulations.};
        \node (Tendons) [process, below of = Reinforcement] {\textbf{Tendons}\\
        Choose tendon amount and distribution based on those used in similar bridge designs.};
        \node (Calculations) [process, below of = Tendons] {\textbf{Calculations}\\
        Perform numerical simulations and calculate the utilization ratio of all relevant constraints.};
        \node (Res) [decision2, below of = Calculations] {Constraints violated \\ or potential improvements?};
        \node (Finished) [start, below of = Res] {\textbf{Design complete}\\ 
            Satisfactory design variables have been determined.};
        \node (Revise) [process, left of = Calculations, xshift = -5.0cm] {\textbf{Adjustments}\\
        Determine the tendon and/or reinforcement quantities to increase to ensure that constraints are not violated, or quantities that can be reduced while still satisfying constraints.};
  
        \draw [arrow] (Initial) -- (GirderType);
        \draw [arrow] (GirderType) -- (GirderHeight);
        \draw [arrow] (GirderHeight) -- (Reinforcement);
        \draw [arrow] (Reinforcement) -- (Tendons);
        \draw [arrow] (Tendons) -- (Calculations);
        \draw [arrow] (Calculations) -- (Res);
        \draw [arrow] (Res) -- node[anchor=east] {No} (Finished);
        \draw [line] (Res.west) -- node[anchor=south] {Yes} ($(Revise.north |- Res.west)$);
        \draw [arrow] ($(Revise.north |- Res.east)$) -- (Revise.south);
        \draw [arrow] (Revise) -- (Calculations);

        \end{tikzpicture}
    \caption{Traditional bridge girder design process}
    \label{fig:ROT_Process}
\end{figure}

As described in the introduction, the traditional method of bridge design is based upon expert opinion and best guesses or rules of thumb which is then followed by multiple iterations to achieve the final design. The traditional design process for a concrete road bridge is illustrated in Figure~\ref{fig:ROT_Process}. The process is based on the authors' experience from working several years in the bridge design industry. The process follows a set of consecutive steps in which one considers one variable or component at a time. When an initial design has been determined, simulations are performed and the utilization ratio of the relevant constraints is calculated. The designer must then evaluate whether certain areas need additional reinforcement or tendons to meet the constraints, or if this can be reduced in certain areas. Decreasing the amount of materials is considered beneficial as it reduces the cost of the structure.

This process was performed for the bridge girder presented in the case study of Section~\ref{Ch. Case-study}. A design was determined in which the minimum reinforcement required by the Eurocodes and national regulations was applied, and the tendon quantity was reduced until no further reduction was possible which was governed by the constraint \emph{Concrete compression at tendons} of Table \ref{table:Constraints}. This design is abbreviated as ROT in the results and serves as a benchmark for the MOO process.

\section{Results}
\subsection{Comparing PTMOO to Benchmark design}
\begin{figure}[H]
    \centering
    \includegraphics[width=0.90\textwidth]{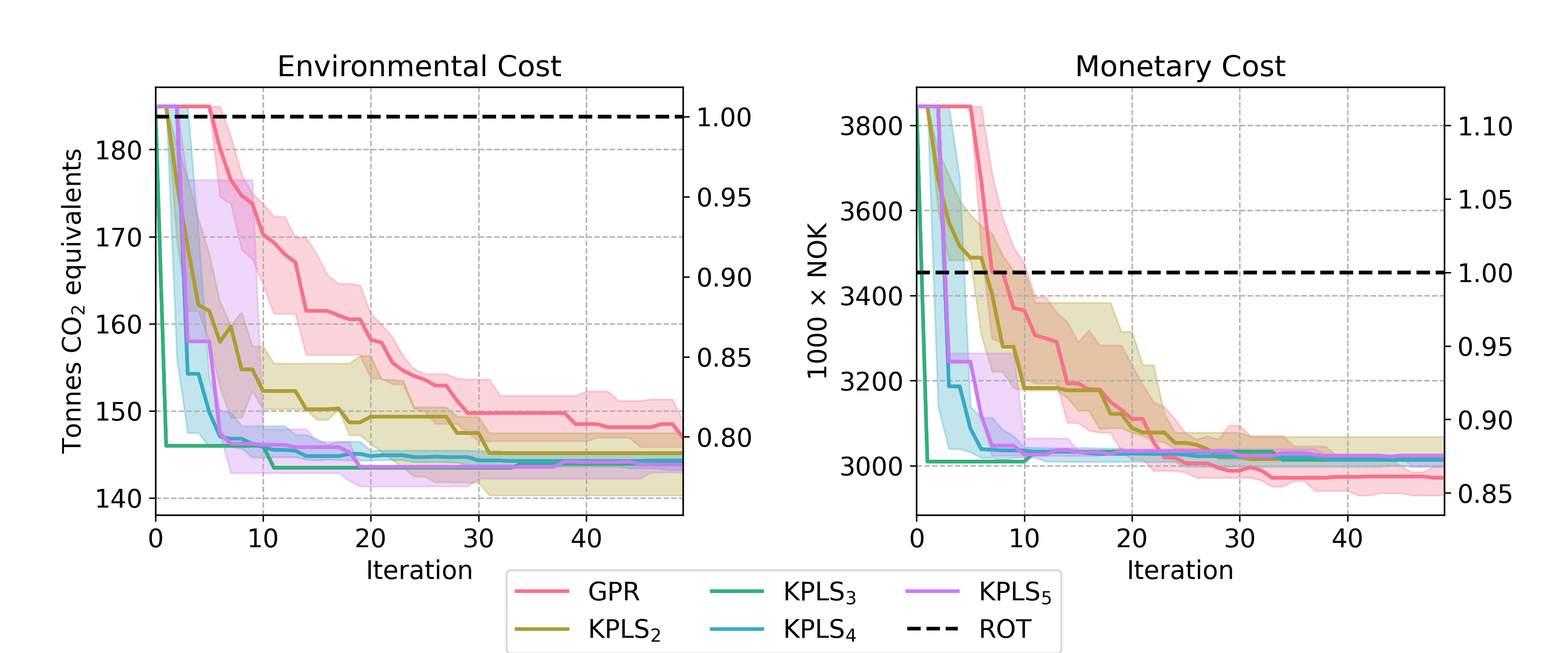}
    \caption{Environmental and monetary cost of the preferred design $\textbf{x}_N^*$ for the PTMOO. Solid line represents the median and the shaded area the 50 \% confidence interval of the 20 repetitions.}
    \label{fig:ObjectiveResults}
\end{figure}

The environmental and monetary cost of the preferred design $\textbf{x}^*_N$ according to Eq. \eqref{Eq. XN_opt}, at each iteration of the PTMOO process is shown in Figure~\ref{fig:ObjectiveResults}. The figure shows the median of the objective functions $\boldsymbol{f}(\textbf{x}_N^*)$ in each iteration with solid lines, and the 50 \% confidence interval as the corresponding shaded area. The figure shows that the PTMOO have an environmental cost which is about 20~\% lower than the ROT, and simultaneously a monetary cost that is between 10~\% and 15~\% lower. The figure also shows that the PTMOO using KPLS converge faster, i.e. reaching a plateau at fewer iterations, than the method that utilizes only POD with full GPR. The KPLS reduces the number of hyperparameters, and since there are limited observations at the start of the optimization process, it seems to be more capable of providing accurate predictions at the early phase. The full GPR appears to converge to a design with a slightly higher environmental cost, albeit slightly lower monetary cost. In the case study, the KPLS with three weight components converge the fastest, while increasing the number of weights beyond three results in a slightly lower convergence rate. All KPLS methods seem to converge to approximately the same optimal environmental and monetary costs.

In Table \ref{table:OptimumVariables}, the variable combination of the preferred design at the last iteration $\textbf{x}^*_{50}$ for ROT, GPR and $\text{KPLS}_5$ is shown. For GPR and $\text{KPLS}_5$ the mean $\pm$ one standard deviation for the 20 evaluations is shown. The table shows that the ROT has a higher girder section with less tendon area, and a higher reinforcement area in the upper section than the PTMOO methods. The minimum reinforcement requirement of the Eurocodes and national regulations is not considered in the PTMOO processes, making it possible that the results do not comply with these requirements.  In practice, such requirements can be satisfied through minor post-optimization adjustments, which are not expected to significantly alter the trade-off behavior, cost or performance of the designs. Since the minimum reinforcement has a linear relationship with the section height, the lower section height in the PTMOO process will result in a lower minimum reinforcement compared to the ROT. It should also be noted that in practice, the reinforcement and tendon amounts must be according to the suppliers' available dimensions.

\begin{table}[h!]
    \centering
    \small
    \caption{Variable combination of the preferred design at the last iteration $\textbf{x}^*_{50}$ of the PTMOO}
    \label{table:OptimumVariables}
    \begin{tabular}{ l l c r r@{\hspace{2pt}}c@{\hspace{2pt}}r r@{\hspace{2pt}}c@{\hspace{2pt}}r}
    \textbf{Variable} & & \textbf{Unit} &\multicolumn{1}{c}{\textbf{ROT}} &\multicolumn{3}{c}{\textbf{GPR}$^*$} &\multicolumn{3}{c}{${\textbf{KPLS}_5}^*$} \\
    \hline
    Tendon strand section area &   &   &\\
    \hspace{2em} Main span & ($A_{TsM}$) & [$\text{mm}^2$] & 127.8 & 130.5 & $\pm$ & 12.8 & 139.8 & $\pm$ & 1.6\\
    \hspace{2em} Side spans & ($A_{TsS}$)& [$\text{mm}^2$] &  62.2 & 74.1 & $\pm$ & 15.0 & 76.0 & $\pm$ & 1.3\\
    Tendon inflection point - Side spans & ($\alpha_{TsS}$) & [-] & 0.60 & 0.54 & $\pm$ & 0.04 & 0.64 & $\pm$ & 0.02\\
    Girder cross section height & ($H$)  & [m] &  1.50 & 1.10 & $\pm$ & 0.10 & 1.01 & $\pm$ & 0.01\\
    Reinforcement area &   &   &   \\
    \hspace{2em} Upper section - Part 1 & ($A_{T1}$) & [$\text{mm}^2$] & 15 134 & 4 366 & $\pm$ & 1 613 & 6 793 & $\pm$ & 74\\
    \hspace{2em} Upper section - Part 2 & ($A_{T2}$) & [$\text{mm}^2$] & 12 382 & 6 491 & $\pm$ & 2 407 & 8 792 & $\pm$ & 360\\
    \hspace{2em} Upper section - Part 3 & ($A_{T3}$) & [$\text{mm}^2$] & 16 234 & 4 908 & $\pm$ & 1 464 & 5 745 & $\pm$ & 131\\
    \hspace{2em} Lower section - Part 1 & ($A_{B1}$) & [$\text{mm}^2$] & 5 928 & 5 223 & $\pm$ & 2 175 & 5 454 & $\pm$ & 527\\
    \hspace{2em} Lower section - Part 2 & ($A_{B2}$) & [$\text{mm}^2$] & 5 928 & 6 339 & $\pm$ & 2 539 & 7 197 & $\pm$ & 52 \\
    \hspace{2em} Lower section - Part 3 & ($A_{B3}$) & [$\text{mm}^2$] & 5 928 & 6 538 & $\pm$ & 1 647 & 8 120 & $\pm$ & 153\\
    \hspace{2em} Lower section - Part 4 & ($A_{B4}$) & [$\text{mm}^2$] & 5 928 & 8 259 & $\pm$ & 1 901 & 9 101 & $\pm$ & 148\\
    Reinforcement distribution length &   &    &  \\
    \hspace{2em} Upper section - Part 1 & ($\alpha_{T1}$) & [-] & 0.85 & 0.81 & $\pm$ & 0.09 & 0.76 & $\pm$ & 0.02\\
    \hspace{2em} Upper section - Part 3 & ($\alpha_{T3}$) & [-] & 0.70 & 0.53 & $\pm$ & 0.32 & 0.89 & $\pm$ & 0.04\\
    \hspace{2em} Lower section - Part 2 & ($\alpha_{B2}$) & [-] & 0.50 & 0.66 & $\pm$ & 0.20 & 0.64 & $\pm$ & 0.02\\
    \hspace{2em} Lower section - Part 4 & ($\alpha_{B4}$) & [-] & 0.50 & 0.61 & $\pm$ & 0.19 & 0.63 & $\pm$ & 0.02\\
    \hline
    \end{tabular}
    {\footnotesize * Values are given as mean $\pm$ one standard deviation.}
\end{table}

\begin{figure}[H]
    \centering
    \includegraphics[width=0.45\textwidth]{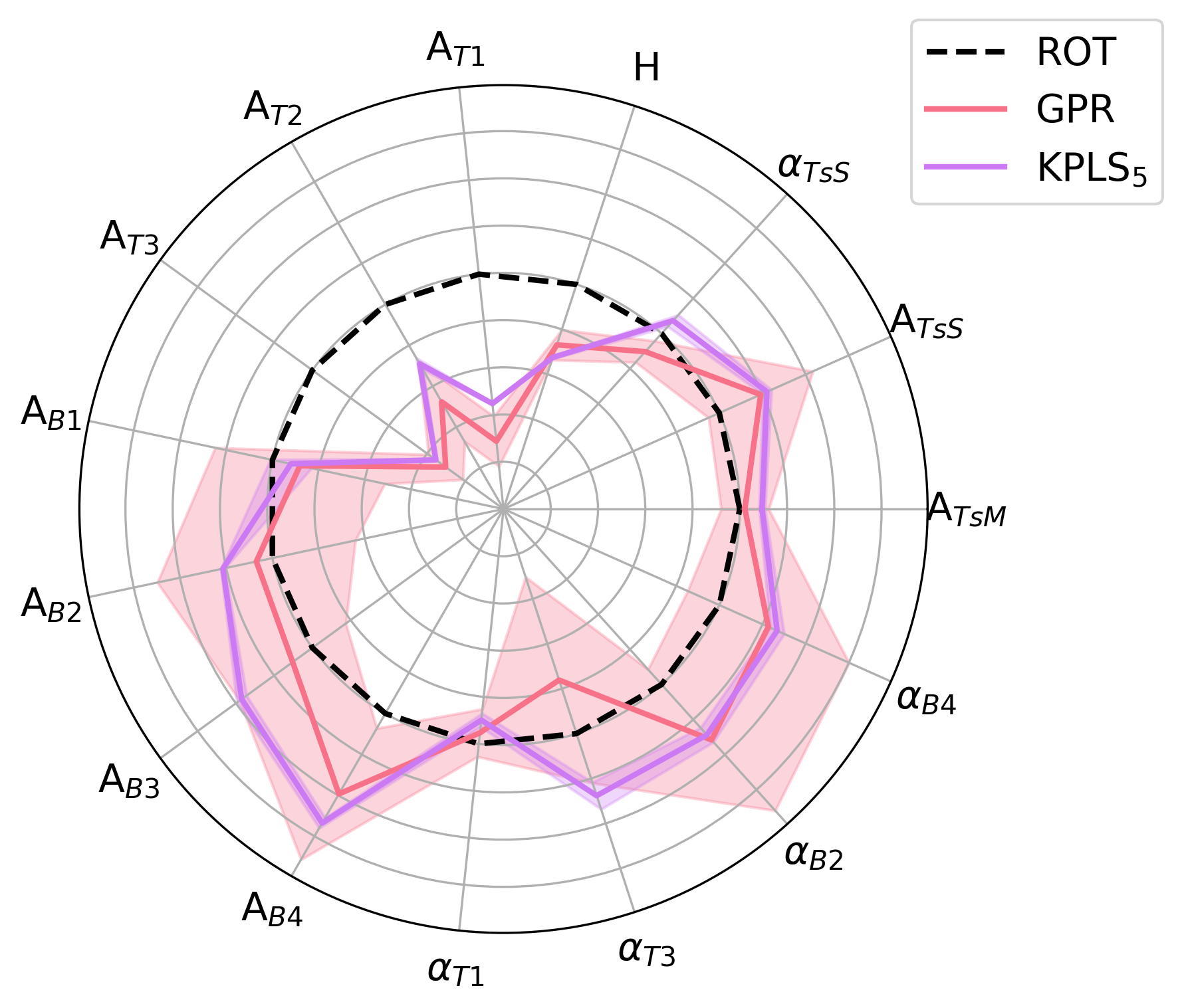}
    \caption{Variable combination of the preferred design at the last iteration $\textbf{x}^*_{50}$ of the PTMOO relative to ROT. The solid
lines show the mean value and the shaded areas are the mean $\pm$ one standard deviation.}
    \label{fig:BestValuesRadarPlot}
\end{figure}

In Table \ref{table:OptimumVariables} we also note that the GPR has a significantly higher standard deviation than the $\text{KPLS}_5$ for the preferred feasible design variables, which is also apparent in the wider confidence interval of Figure \ref{fig:ObjectiveResults}. In Figure \ref{fig:BestValuesRadarPlot} the variable combination of the preferred design at the last iteration $\textbf{x}^*_{50}$ relative to ROT is illustrated. The solid lines show the mean value and the shaded areas are the mean $\pm$ one standard deviation. The likelihood function of Eq. \eqref{Eq:marglikelihood}, which is used to determine the hyperparameters, is often multimodal. Because GPR must learn many more hyperparameters, its training is more susceptible to getting stuck in local optima, which can degrade prediction accuracy. In contrast, $\text{KPLS}_5$ is limited to just five hyperparameters, greatly reducing its flexibility and thus its vulnerability to suboptimal fits. This difference in model complexity likely accounts for the larger standard deviation observed in the preferred feasible design variables when using GPR.

\begin{figure}[H]
    \centering
    \includegraphics[width=0.90\textwidth]{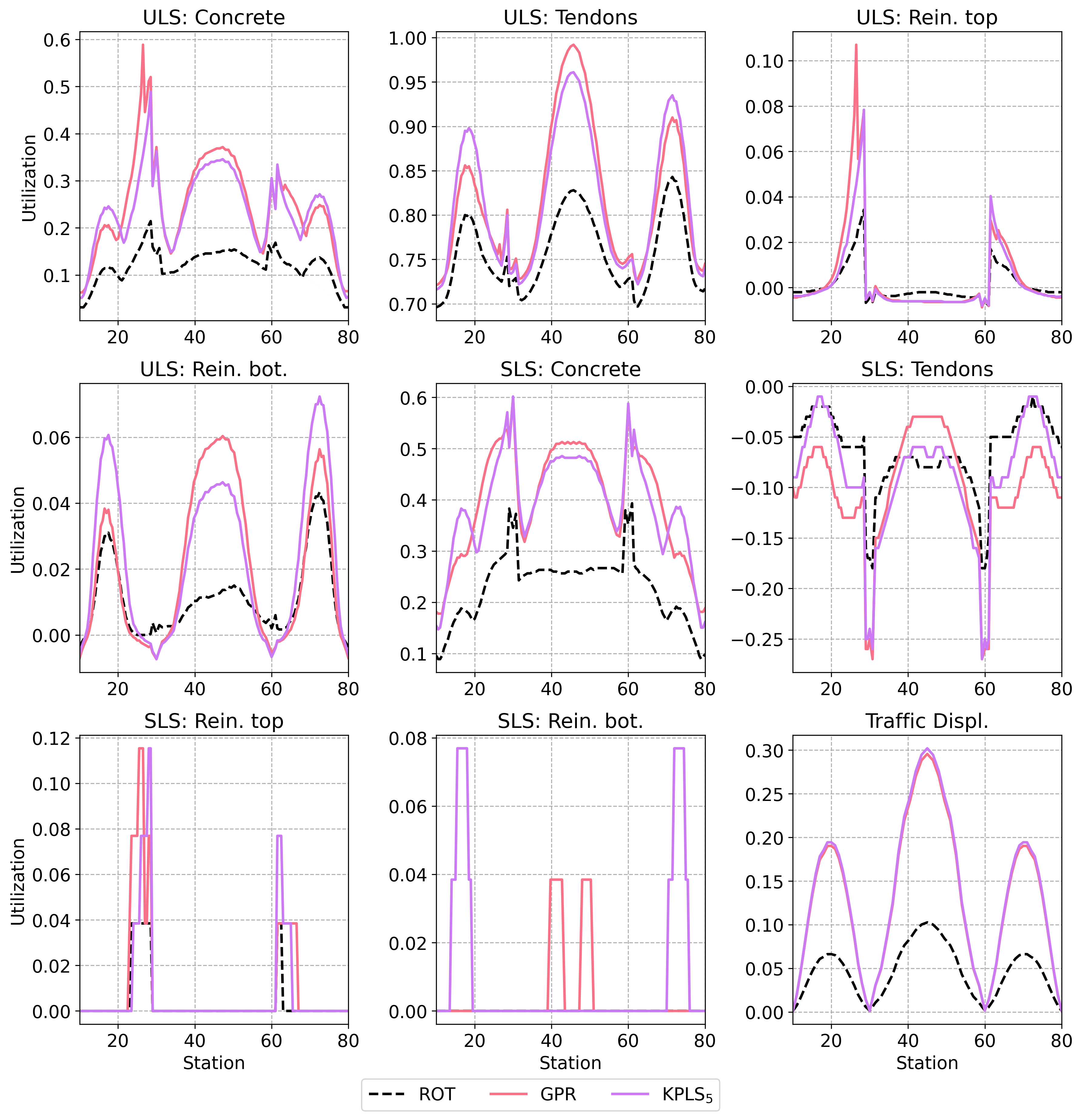}
    \caption{Constraints along the bridge for the preferred solution at final iteration for ROT, GPR and $\text{KPLS}_5$ using PTMOO. SLS and ULS are the limit states described in Section \ref{LoadCombs}. "Rein." is normal reiforcement, "bot." is at the bottom of the girder and "top" is at the top of the girder. Limit is 1.0 for all constraints except "SLS: Tendons" where it is 0.0.}
    \label{fig:ConstrainResults}
\end{figure}

Figure \ref{fig:ConstrainResults} presents the utilization ratio of the constraints for the preferred solution along the bridge, referred to as the \emph{station} in the figure, at the final iteration of the PTMOO process. The results are shown for a randomly selected case from the 20 repetitions, comparing the process using $\text{KPLS}_5$, GPR, and the benchmark ROT. Constraints are satisfied as long as they are below 1.0, or 0.0 for the \emph{SLS: Tendons} case. As the figure shows, all constraints are satisfied. What is interesting to note is that the $\text{KPLS}_5$ and the GPR have relatively similar utilizations, sometimes the $\text{KPLS}_5$ is slightly higher and for others it is the opposite. In all cases, the ROT has the lowest utilization, except for the \emph{SLS: Tendons}, where utilization levels are relatively similar. This constraint governed the tendon amount in the ROT design process and appears to be quite active in the PTMOO processes as well. Higher utilization is not inherently disadvantageous, as long as it remains within the allowable limit. In fact, it appears that increasing the overall level of utilization improves the efficiency of the load bearing capacity of the structure, ultimately reducing both environmental and financial costs. The figure demonstrates that considering multiple variables simultaneously, as is done in the PTMOO processes, can lead to greater structural utilization, which further contributes to cost and environmental benefits.

\subsection{Comparing cEHVI to PTMOO}
\begin{figure}[H]
    \centering
    \includegraphics[width=0.9\linewidth]{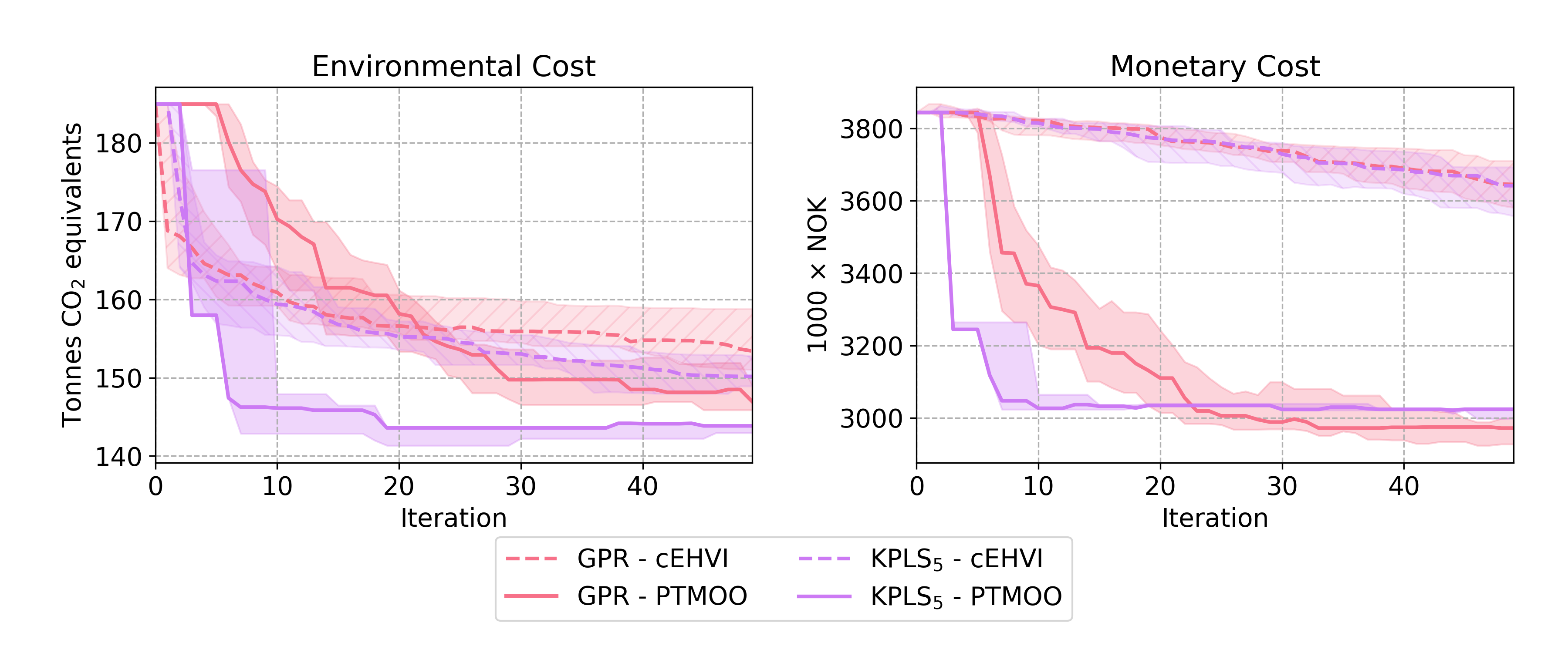}
    \caption{cEHVI compared to PTMOO. Environmental and monetary cost of the preferred design $\textbf{x}^*_{N}$ at each iteration according to Eq. \eqref{Eq. XN_opt}. Solid line represents the median and the shaded area the 50 \% confidence interval of the 20 repetitions for the PTMOO. The dashed line and the hatched shaded area represent the corresponding results for the cEHVI approach.}
    \label{fig:cEHVI_vs_PTMOO}
\end{figure}

Figure \ref{fig:cEHVI_vs_PTMOO} depicts, at each iteration and in accordance with Eq. \eqref{Eq. XN_opt}, the monetary and environmental cost of the preferred design $\textbf{x}^*_{N}$ found by both optimization schemes. The solid curves show the PTMOO results, with the shaded region indicating the 50 \% confidence interval. The dashed curves show the cEHVI results, with the hatched region indicating the corresponding 50 \% confidence interval. Figure \ref{fig:cEHVI_vs_PTMOO} demonstrates that, over 50 iterations, PTMOO identifies a design incurring both lower monetary and environmental costs. Although $\text{KPLS}_5$ appears to plateau after roughly 20 iterations, cEHVI is still improving at iteration 50. This slower convergence is expected. cEHVI explores multiple feasible solutions under arbitrary objective trade‐offs, whereas PTMOO targets a single solution with its trade‐off fixed by Eq.\eqref{Eq. XN_opt}. 

\subsection{Comparing GBO to PTMOO}
\begin{figure}[H]
    \centering
    \includegraphics[width=0.9\linewidth]{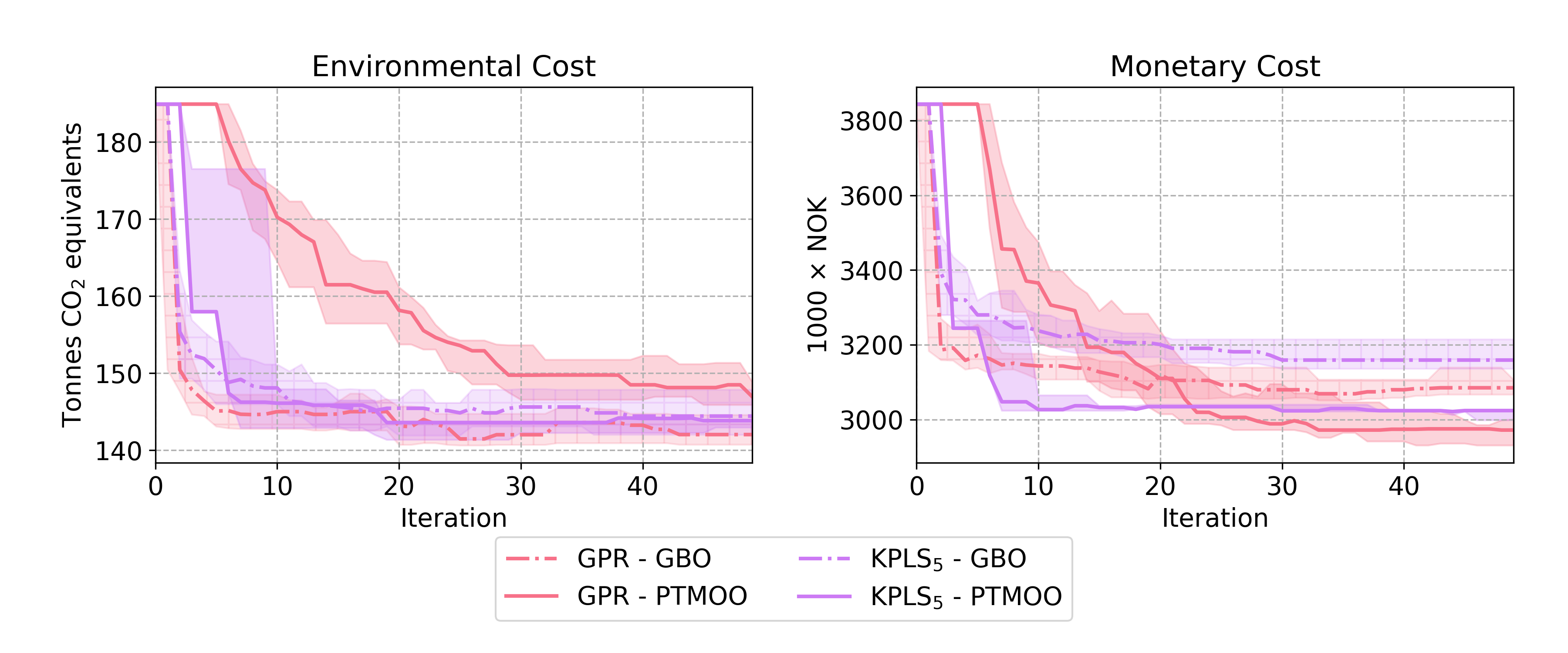}
    \caption{GBO compared to PTMOO. Environmental and monetary cost of the preferred design $\textbf{x}^*_{N}$ at each iteration according to Eq. \eqref{Eq. XN_opt}. Solid line represents the median and the shaded area the 50 \% confidence interval of the 20 repetitions for the PTMOO. The dot-dashed line and the hatched shaded area represent the corresponding results for the GBO approach.}
    \label{fig:GB_vs_PTMOO}
\end{figure}

In Figure \ref{fig:GB_vs_PTMOO}, the results of PTMOO are compared with GBO, showing the environmental and monetary costs of the preferred design $\textbf{x}^*_{N}$ as defined in Eq. \eqref{Eq. XN_opt}. The figure reports the median performance, with solid lines for PTMOO and dash-dotted lines for GBO, while the 50 \% confidence intervals are indicated by shaded regions for the PTMOO and hatched regions for the GBO.

GBO without KPLS (GPR-GBO) outperforms the convergence rate of PTMOO without KPLS (GPR-TMOO), converging more rapidly to designs with lower environmental costs, albeit a higher monetary cost. When KPLS is included, the KPLS appear to have a degrading effect on the optimization process for the GBO approach, which is opposite to what is observed for the PTMOO. The GBO tends to converge to slightly higher monetary costs than the PTMOO. The faster convergence of GBO reflects its reliance on gradient information to efficiently exploit promising descent directions, whereas PTMOO maintains a broader exploration of the design space, enabling the identification of better optima. The use of KPLS appears to improve the PTMOO by simplifying the design space through dimensionality reduction, while it appears to limit the search space of the GBO. 

\section{Conclusion}
A multiobjective constrained Bayesian optimization process with a preferred trade-off for the objectives has been presented and applied to a 3-span post-tensioned concrete road bridge girder. Fifteen design variables were considered in the optimization process to minimize both the environmental and monetary cost of the structure, while also fulfilling nine constraints. Optimization was performed using EI$_c$ Bayesian acquisition functions, where the improvement was measured relative to the current preferred feasible solution. Both constraints and objectives were approximated using GPR. Due to the high dimensionality of the constraints, which were derived from numerical simulations, they were approximated by using POD with GPR to approximate unobserved design variable combinations. In addition, the use of KPLS was explored to assess its potential to enhance the effectiveness of the optimization process.

As a benchmark for the case-study, design variables and corresponding environmental and monetary costs were determined using a traditional bridge girder design process. The results showed that multi-objective optimization methods with a preferred trade-off for objectives were able to reduce the environmental cost with about 20~\% and the monetary cost with between 10~\% and 15~\% while complying to all constraints, when compared to the benchmark. The results also showed that applying KPLS can result in a faster convergence, i.e. fewer number of iterations before plateauing, than GPR with a full number of kernel parameters. For the case study, a minimum of three weight components were necessary to achieve this. The KPLS and GPR with a full number of kernel parameters converged to design variables that resulted in about the same monetary and environmental costs.

The MOO routine with a preferred trade-off was compared to the commonly used cEHVI MOO routine, where the trade-off is arbitrary. The results showed that the approach with a preferred trade-off was able to obtain a solution where both the environmental and monetary cost were lower than that achieved using cEHVI with the same number of iterations. This is as expected, as the cEHVI expends effort exploring the full set of non-dominated solutions. For the cEHVI, the objectives were normalized and bounded by a reference point encompassing the full possible Pareto front. A more aggressive approach, if the extremities of the Pareto front is not of interest, would be to use the preferred trade-off solution, e.g. Eq. \eqref{Eq. XN_opt}, as a reference point.

The MOO method presented in this article is applicable to any optimization problem where data is expensive or time consuming to generate, resulting in a limited amount of available data, and a moderate number of design variables are considered. These are conditions that are common in many engineering design problems that involve numerical simulations. The article has also shown that considering multiple variables simultaneously can lead to higher structural utilization, ultimately resulting in lower environmental and monetary costs. A bottleneck for the applicability of the presented MOO is generating enough observations to make accurate predictions. Since simulations can be time-consuming, reducing their number would be beneficial. Future research will explore the potential of using deep Gaussian process regressors for the MOO method to determine whether they can improve prediction accuracy with less data.

\section*{Statements and Declarations}
\subsection*{Acknowledgments}
A special thanks to assistant professor Frederico Afonso at Instituto Superior Técnico (IST) - University of Lisbon, for his thoughts and advises on the multi-objective optimization routine.

\subsection*{Funding}
This work is funded by the Research Council of Norway [project number 332126], and by Dr Ing A Aas-Jakobsen AS.

\subsection*{Data Availability}
Results from the finite element analyses derived from the case study of Section \ref{Ch. Case-study} can be made available upon request by e-mailing the corresponding author.

\subsection*{Replication of results}
The method used for the optimization process was performed as described in the article. For the purpose of a realistic industrial application, the bridge in the case-study was analyzed using the in-house software NovaFrame and is therefore subject to strict confidentiality.

\subsection*{Ethics approval and Consent to participate }
Not applicable.

\subsection*{Author contributions}
The authors’ contribution is as follows: Heine Røstum was responsible for the conceptualization, methodology, formal analyses, data curation, software writing and original draft work. Joseph Morlier was responsible for supervision, conceptualization and methodology. Sebastien Gros and Ketil Aas-Jakobsen were responsible for the supervision and project administration.

\subsection*{Conflict of Interest}
On behalf of all authors, the corresponding author states that there is no conflict of interest.

\bibliography{Bibliography}% common bib file


\begin{thebibliography}{45}
\ifx \bisbn   \undefined \def \bisbn  #1{ISBN #1}\fi
\ifx \binits  \undefined \def \binits#1{#1}\fi
\ifx \bauthor  \undefined \def \bauthor#1{#1}\fi
\ifx \batitle  \undefined \def \batitle#1{#1}\fi
\ifx \bjtitle  \undefined \def \bjtitle#1{#1}\fi
\ifx \bvolume  \undefined \def \bvolume#1{\textbf{#1}}\fi
\ifx \byear  \undefined \def \byear#1{#1}\fi
\ifx \bissue  \undefined \def \bissue#1{#1}\fi
\ifx \bfpage  \undefined \def \bfpage#1{#1}\fi
\ifx \blpage  \undefined \def \blpage #1{#1}\fi
\ifx \burl  \undefined \def \burl#1{\textsf{#1}}\fi
\ifx \doiurl  \undefined \def \doiurl#1{\url{https://doi.org/#1}}\fi
\ifx \betal  \undefined \def \betal{\textit{et al.}}\fi
\ifx \binstitute  \undefined \def \binstitute#1{#1}\fi
\ifx \binstitutionaled  \undefined \def \binstitutionaled#1{#1}\fi
\ifx \bctitle  \undefined \def \bctitle#1{#1}\fi
\ifx \beditor  \undefined \def \beditor#1{#1}\fi
\ifx \bpublisher  \undefined \def \bpublisher#1{#1}\fi
\ifx \bbtitle  \undefined \def \bbtitle#1{#1}\fi
\ifx \bedition  \undefined \def \bedition#1{#1}\fi
\ifx \bseriesno  \undefined \def \bseriesno#1{#1}\fi
\ifx \blocation  \undefined \def \blocation#1{#1}\fi
\ifx \bsertitle  \undefined \def \bsertitle#1{#1}\fi
\ifx \bsnm \undefined \def \bsnm#1{#1}\fi
\ifx \bsuffix \undefined \def \bsuffix#1{#1}\fi
\ifx \bparticle \undefined \def \bparticle#1{#1}\fi
\ifx \barticle \undefined \def \barticle#1{#1}\fi
\bibcommenthead
\ifx \bconfdate \undefined \def \bconfdate #1{#1}\fi
\ifx \botherref \undefined \def \botherref #1{#1}\fi
\ifx \url \undefined \def \url#1{\textsf{#1}}\fi
\ifx \bchapter \undefined \def \bchapter#1{#1}\fi
\ifx \bbook \undefined \def \bbook#1{#1}\fi
\ifx \bcomment \undefined \def \bcomment#1{#1}\fi
\ifx \oauthor \undefined \def \oauthor#1{#1}\fi
\ifx \citeauthoryear \undefined \def \citeauthoryear#1{#1}\fi
\ifx \endbibitem  \undefined \def \endbibitem {}\fi
\ifx \bconflocation  \undefined \def \bconflocation#1{#1}\fi
\ifx \arxivurl  \undefined \def \arxivurl#1{\textsf{#1}}\fi
\csname PreBibitemsHook\endcsname

%%% 1
\bibitem[\protect\citeauthoryear{Abadi
  et~al.}{2015}]{tensorflow2015-whitepaper}
\begin{botherref}
\oauthor{\bsnm{Abadi}, \binits{M.}},
\oauthor{\bsnm{Agarwal}, \binits{A.}},
\oauthor{\bsnm{Barham}, \binits{P.}},
\oauthor{\bsnm{Brevdo}, \binits{E.}},
\oauthor{\bsnm{Chen}, \binits{Z.}},
\oauthor{\bsnm{Citro}, \binits{C.}},
\oauthor{\bsnm{Corrado}, \binits{G.S.}},
\oauthor{\bsnm{Davis}, \binits{A.}},
\oauthor{\bsnm{Dean}, \binits{J.}},
\oauthor{\bsnm{Devin}, \binits{M.}},
\oauthor{\bsnm{Ghemawat}, \binits{S.}},
\oauthor{\bsnm{Goodfellow}, \binits{I.}},
\oauthor{\bsnm{Harp}, \binits{A.}},
\oauthor{\bsnm{Irving}, \binits{G.}},
\oauthor{\bsnm{Isard}, \binits{M.}},
\oauthor{\bsnm{Jia}, \binits{Y.}},
\oauthor{\bsnm{Jozefowicz}, \binits{R.}},
\oauthor{\bsnm{Kaiser}, \binits{L.}},
\oauthor{\bsnm{Kudlur}, \binits{M.}},
\oauthor{\bsnm{Levenberg}, \binits{J.}},
\oauthor{\bsnm{Mané}, \binits{D.}},
\oauthor{\bsnm{Monga}, \binits{R.}},
\oauthor{\bsnm{Moore}, \binits{S.}},
\oauthor{\bsnm{Murray}, \binits{D.}},
\oauthor{\bsnm{Olah}, \binits{C.}},
\oauthor{\bsnm{Schuster}, \binits{M.}},
\oauthor{\bsnm{Shlens}, \binits{J.}},
\oauthor{\bsnm{Steiner}, \binits{B.}},
\oauthor{\bsnm{Sutskever}, \binits{I.}},
\oauthor{\bsnm{Talwar}, \binits{K.}},
\oauthor{\bsnm{Tucker}, \binits{P.}},
\oauthor{\bsnm{Vanhoucke}, \binits{V.}},
\oauthor{\bsnm{Vasudevan}, \binits{V.}},
\oauthor{\bsnm{Viégas}, \binits{F.}},
\oauthor{\bsnm{Vinyals}, \binits{O.}},
\oauthor{\bsnm{Warden}, \binits{P.}},
\oauthor{\bsnm{Wattenberg}, \binits{M.}},
\oauthor{\bsnm{Wicke}, \binits{M.}},
\oauthor{\bsnm{Yu}, \binits{Y.}},
\oauthor{\bsnm{Zheng}, \binits{X.}}:
TensorFlow: Large-Scale Machine Learning on Heterogeneous Systems.
Software available from tensorflow.org
(2015).
\url{https://www.tensorflow.org/}
\end{botherref}
\endbibitem

%%% 2
\bibitem[\protect\citeauthoryear{Amsallem and
  Farhat}{2012}]{amsallem2012stabilization}
\begin{barticle}
\bauthor{\bsnm{Amsallem}, \binits{D.}},
\bauthor{\bsnm{Farhat}, \binits{C.}}:
\batitle{Stabilization of projection-based reduced-order models}.
\bjtitle{International Journal for Numerical Methods in Engineering}
\bvolume{91}(\bissue{4}),
\bfpage{358}--\blpage{377}
(\byear{2012})
\end{barticle}
\endbibitem

%%% 3
\bibitem[\protect\citeauthoryear{Bouhlel et~al.}{2016}]{bouhlel2016improving}
\begin{barticle}
\bauthor{\bsnm{Bouhlel}, \binits{M.A.}},
\bauthor{\bsnm{Bartoli}, \binits{N.}},
\bauthor{\bsnm{Otsmane}, \binits{A.}},
\bauthor{\bsnm{Morlier}, \binits{J.}}:
\batitle{Improving kriging surrogates of high-dimensional design models by
  partial least squares dimension reduction}.
\bjtitle{Structural and Multidisciplinary Optimization}
\bvolume{53},
\bfpage{935}--\blpage{952}
(\byear{2016})
\end{barticle}
\endbibitem

%%% 4
\bibitem[\protect\citeauthoryear{{Blank} and {Deb}}{2020}]{pymoo}
\begin{barticle}
\bauthor{\bsnm{{Blank}}, \binits{J.}},
\bauthor{\bsnm{{Deb}}, \binits{K.}}:
\batitle{pymoo: Multi-objective optimization in python}.
\bjtitle{IEEE Access}
\bvolume{8},
\bfpage{89497}--\blpage{89509}
(\byear{2020})
\end{barticle}
\endbibitem

%%% 5
\bibitem[\protect\citeauthoryear{Brunton and Kutz}{2022}]{brunton2022data}
\begin{bbook}
\bauthor{\bsnm{Brunton}, \binits{S.L.}},
\bauthor{\bsnm{Kutz}, \binits{J.N.}}:
\bbtitle{Data-driven Science and Engineering: Machine Learning, Dynamical
  Systems, and Control}.
\bpublisher{Cambridge University Press},
\blocation{Cambridge, England}
(\byear{2022})
\end{bbook}
\endbibitem

%%% 6
\bibitem[\protect\citeauthoryear{Buljak and Maier}{2011}]{buljak2011proper}
\begin{barticle}
\bauthor{\bsnm{Buljak}, \binits{V.}},
\bauthor{\bsnm{Maier}, \binits{G.}}:
\batitle{Proper orthogonal decomposition and radial basis functions in material
  characterization based on instrumented indentation}.
\bjtitle{Engineering Structures}
\bvolume{33}(\bissue{2}),
\bfpage{492}--\blpage{501}
(\byear{2011})
\end{barticle}
\endbibitem

%%% 7
\bibitem[\protect\citeauthoryear{Chiplunkar
  et~al.}{2018}]{chiplunkar2018gaussian}
\begin{bchapter}
\bauthor{\bsnm{Chiplunkar}, \binits{A.}},
\bauthor{\bsnm{Bosco}, \binits{E.}},
\bauthor{\bsnm{Morlier}, \binits{J.}}:
\bctitle{Gaussian process for aerodynamic pressures prediction in fast fluid
  structure interaction simulations}.
In: \bbtitle{Advances in Structural and Multidisciplinary Optimization:
  Proceedings of the 12th World Congress of Structural and Multidisciplinary
  Optimization (WCSMO12) 12},
pp. \bfpage{221}--\blpage{233}
(\byear{2018}).
\bcomment{Springer}
\end{bchapter}
\endbibitem

%%% 8
\bibitem[\protect\citeauthoryear{Chatterjee}{2000}]{chatterjee2000introduction}
\begin{botherref}
\oauthor{\bsnm{Chatterjee}, \binits{A.}}:
An introduction to the proper orthogonal decomposition.
Current science,
808--817
(2000)
\end{botherref}
\endbibitem

%%% 9
\bibitem[\protect\citeauthoryear{Daulton
  et~al.}{2020}]{daulton2020differentiable}
\begin{barticle}
\bauthor{\bsnm{Daulton}, \binits{S.}},
\bauthor{\bsnm{Balandat}, \binits{M.}},
\bauthor{\bsnm{Bakshy}, \binits{E.}}:
\batitle{Differentiable expected hypervolume improvement for parallel
  multi-objective bayesian optimization}.
\bjtitle{Advances in Neural Information Processing Systems}
\bvolume{33},
\bfpage{9851}--\blpage{9864}
(\byear{2020})
\end{barticle}
\endbibitem

%%% 10
\bibitem[\protect\citeauthoryear{de~Gooijer et~al.}{2021}]{de2021evaluation}
\begin{barticle}
\bauthor{\bsnm{Gooijer}, \binits{B.M.}},
\bauthor{\bsnm{Havinga}, \binits{J.}},
\bauthor{\bsnm{Geijselaers}, \binits{H.J.}},
\bauthor{\bsnm{Boogaard}, \binits{A.H.}}:
\batitle{Evaluation of pod based surrogate models of fields resulting from
  nonlinear fem simulations}.
\bjtitle{Advanced Modeling and Simulation in Engineering Sciences}
\bvolume{8}(\bissue{1}),
\bfpage{25}
(\byear{2021})
\end{barticle}
\endbibitem

%%% 11
\bibitem[\protect\citeauthoryear{Drakoulas et~al.}{2024}]{drakoulas2024physics}
\begin{barticle}
\bauthor{\bsnm{Drakoulas}, \binits{G.}},
\bauthor{\bsnm{Gortsas}, \binits{T.V.}},
\bauthor{\bsnm{Polyzos}, \binits{D.}}:
\batitle{Physics-based reduced order modeling for uncertainty quantification of
  guided wave propagation using bayesian optimization}.
\bjtitle{Engineering Applications of Artificial Intelligence}
\bvolume{133},
\bfpage{108531}
(\byear{2024})
\end{barticle}
\endbibitem

%%% 12
\bibitem[\protect\citeauthoryear{Deb et~al.}{2002}]{deb2002fast}
\begin{barticle}
\bauthor{\bsnm{Deb}, \binits{K.}},
\bauthor{\bsnm{Pratap}, \binits{A.}},
\bauthor{\bsnm{Agarwal}, \binits{S.}},
\bauthor{\bsnm{Meyarivan}, \binits{T.}}:
\batitle{A fast and elitist multiobjective genetic algorithm: Nsga-ii}.
\bjtitle{IEEE transactions on evolutionary computation}
\bvolume{6}(\bissue{2}),
\bfpage{182}--\blpage{197}
(\byear{2002})
\end{barticle}
\endbibitem

%%% 13
\bibitem[\protect\citeauthoryear{{Dr.\ Ing.\ A. Aas-Jakobsen
  A/S}}{2020}]{NovaFrame6_2}
\begin{botherref}
\oauthor{\bsnm{{Dr.\ Ing.\ A. Aas-Jakobsen A/S}}}:
{NovaFrame}: In-house bridge design software.
Developed by {Dr.\ Ing.\ A. Aas-Jakobsen A/S},
Oslo, Norway
(2020)
\end{botherref}
\endbibitem

%%% 14
\bibitem[\protect\citeauthoryear{Eckart and
  Young}{1936}]{eckart1936approximation}
\begin{barticle}
\bauthor{\bsnm{Eckart}, \binits{C.}},
\bauthor{\bsnm{Young}, \binits{G.}}:
\batitle{The approximation of one matrix by another of lower rank}.
\bjtitle{Psychometrika}
\bvolume{1}(\bissue{3}),
\bfpage{211}--\blpage{218}
(\byear{1936})
\end{barticle}
\endbibitem

%%% 15
\bibitem[\protect\citeauthoryear{Frazier}{2018}]{frazier2018bayesian}
\begin{bchapter}
\bauthor{\bsnm{Frazier}, \binits{P.I.}}:
\bctitle{Bayesian optimization}.
In: \bbtitle{Recent Advances in Optimization and Modeling of Contemporary
  Problems},
pp. \bfpage{255}--\blpage{278}.
\bpublisher{Informs},
\blocation{MD, USA}
(\byear{2018})
\end{bchapter}
\endbibitem

%%% 16
\bibitem[\protect\citeauthoryear{Frazier and Wang}{2016}]{frazier2016bayesian}
\begin{botherref}
\oauthor{\bsnm{Frazier}, \binits{P.I.}},
\oauthor{\bsnm{Wang}, \binits{J.}}:
Bayesian optimization for materials design.
Information science for materials discovery and design,
45--75
(2016)
\end{botherref}
\endbibitem

%%% 17
\bibitem[\protect\citeauthoryear{Guo and Hesthaven}{2018}]{guo2018reduced}
\begin{barticle}
\bauthor{\bsnm{Guo}, \binits{M.}},
\bauthor{\bsnm{Hesthaven}, \binits{J.S.}}:
\batitle{Reduced order modeling for nonlinear structural analysis using
  gaussian process regression}.
\bjtitle{Computer methods in applied mechanics and engineering}
\bvolume{341},
\bfpage{807}--\blpage{826}
(\byear{2018})
\end{barticle}
\endbibitem

%%% 18
\bibitem[\protect\citeauthoryear{Gardner et~al.}{2014}]{gardner2014bayesian}
\begin{bchapter}
\bauthor{\bsnm{Gardner}, \binits{J.R.}},
\bauthor{\bsnm{Kusner}, \binits{M.J.}},
\bauthor{\bsnm{Xu}, \binits{Z.E.}},
\bauthor{\bsnm{Weinberger}, \binits{K.Q.}},
\bauthor{\bsnm{Cunningham}, \binits{J.P.}}:
\bctitle{Bayesian optimization with inequality constraints.}
In: \bbtitle{ICML},
vol. \bseriesno{2014},
pp. \bfpage{937}--\blpage{945}
(\byear{2014})
\end{bchapter}
\endbibitem

%%% 19
\bibitem[\protect\citeauthoryear{Helland}{1988}]{helland1988structure}
\begin{barticle}
\bauthor{\bsnm{Helland}, \binits{I.S.}}:
\batitle{On the structure of partial least squares regression}.
\bjtitle{Communications in statistics-Simulation and Computation}
\bvolume{17}(\bissue{2}),
\bfpage{581}--\blpage{607}
(\byear{1988})
\end{barticle}
\endbibitem

%%% 20
\bibitem[\protect\citeauthoryear{Jeong and Obayashi}{2005}]{jeong2005efficient}
\begin{bchapter}
\bauthor{\bsnm{Jeong}, \binits{S.}},
\bauthor{\bsnm{Obayashi}, \binits{S.}}:
\bctitle{Efficient global optimization (ego) for multi-objective problem and
  data mining}.
In: \bbtitle{2005 IEEE Congress on Evolutionary Computation},
vol. \bseriesno{3},
pp. \bfpage{2138}--\blpage{2145}
(\byear{2005}).
\bcomment{IEEE}
\end{bchapter}
\endbibitem

%%% 21
\bibitem[\protect\citeauthoryear{Jones et~al.}{1998}]{jones1998efficient}
\begin{barticle}
\bauthor{\bsnm{Jones}, \binits{D.R.}},
\bauthor{\bsnm{Schonlau}, \binits{M.}},
\bauthor{\bsnm{Welch}, \binits{W.J.}}:
\batitle{Efficient global optimization of expensive black-box functions}.
\bjtitle{Journal of Global optimization}
\bvolume{13}(\bissue{4}),
\bfpage{455}--\blpage{492}
(\byear{1998})
\end{barticle}
\endbibitem

%%% 22
\bibitem[\protect\citeauthoryear{Knowles}{2005}]{knowles2005parego}
\begin{botherref}
\oauthor{\bsnm{Knowles}, \binits{J.}}:
Parego: A hybrid algorithm with on-line landscape approximation for expensive
  multiobjective optimization problems.
IEEE TRANSACTIONS ON EVOLUTIONARY COMPUTATION
\textbf{10}(1)
(2005)
\end{botherref}
\endbibitem

%%% 23
\bibitem[\protect\citeauthoryear{Lucia et~al.}{2004}]{lucia2004reduced}
\begin{barticle}
\bauthor{\bsnm{Lucia}, \binits{D.J.}},
\bauthor{\bsnm{Beran}, \binits{P.S.}},
\bauthor{\bsnm{Silva}, \binits{W.A.}}:
\batitle{Reduced-order modeling: new approaches for computational physics}.
\bjtitle{Progress in aerospace sciences}
\bvolume{40}(\bissue{1-2}),
\bfpage{51}--\blpage{117}
(\byear{2004})
\end{barticle}
\endbibitem

%%% 24
\bibitem[\protect\citeauthoryear{Liang et~al.}{2002}]{liang2002proper}
\begin{barticle}
\bauthor{\bsnm{Liang}, \binits{Y.}},
\bauthor{\bsnm{Lee}, \binits{H.}},
\bauthor{\bsnm{Lim}, \binits{S.}},
\bauthor{\bsnm{Lin}, \binits{W.}},
\bauthor{\bsnm{Lee}, \binits{K.}},
\bauthor{\bsnm{Wu}, \binits{C.}}:
\batitle{Proper orthogonal decomposition and its applications—part i:
  Theory}.
\bjtitle{Journal of Sound and vibration}
\bvolume{252}(\bissue{3}),
\bfpage{527}--\blpage{544}
(\byear{2002})
\end{barticle}
\endbibitem

%%% 25
\bibitem[\protect\citeauthoryear{McKay et~al.}{1979}]{artLHS}
\begin{barticle}
\bauthor{\bsnm{McKay}, \binits{M.D.}},
\bauthor{\bsnm{Beckman}, \binits{R.J.}},
\bauthor{\bsnm{Conover}, \binits{W.J.}}:
\batitle{A comparison of three methods for selecting values of input variables
  in the analysis of output from a computer code}.
\bjtitle{Technometrics}
\bvolume{21}(\bissue{2}),
\bfpage{239}--\blpage{245}
(\byear{1979})
\end{barticle}
\endbibitem

%%% 26
\bibitem[\protect\citeauthoryear{Martins and
  Ning}{2021}]{martins2021engineering}
\begin{bbook}
\bauthor{\bsnm{Martins}, \binits{J.R.}},
\bauthor{\bsnm{Ning}, \binits{A.}}:
\bbtitle{Engineering Design Optimization}.
\bpublisher{Cambridge University Press},
\blocation{Cambridge, England}
(\byear{2021})
\end{bbook}
\endbibitem

%%% 27
\bibitem[\protect\citeauthoryear{Mathern et~al.}{2021}]{mathern2021multi}
\begin{barticle}
\bauthor{\bsnm{Mathern}, \binits{A.}},
\bauthor{\bsnm{Steinholtz}, \binits{O.S.}},
\bauthor{\bsnm{Sj{\"o}berg}, \binits{A.}},
\bauthor{\bsnm{{\"O}nnheim}, \binits{M.}},
\bauthor{\bsnm{Ek}, \binits{K.}},
\bauthor{\bsnm{Rempling}, \binits{R.}},
\bauthor{\bsnm{Gustavsson}, \binits{E.}},
\bauthor{\bsnm{Jirstrand}, \binits{M.}}:
\batitle{Multi-objective constrained bayesian optimization for structural
  design}.
\bjtitle{Structural and Multidisciplinary Optimization}
\bvolume{63},
\bfpage{689}--\blpage{701}
(\byear{2021})
\end{barticle}
\endbibitem

%%% 28
\bibitem[\protect\citeauthoryear{Negrin
  et~al.}{2023}]{LiteratureReview_MetamodelStructures}
\begin{bchapter}
\bauthor{\bsnm{Negrin}, \binits{I.}},
\bauthor{\bsnm{Kripka}, \binits{M.}},
\bauthor{\bsnm{Yepes}, \binits{V.}}:
\bctitle{Metamodel-assisted design optimization in the field of structural
  engineering: A literature review}.
In: \bbtitle{Structures},
vol. \bseriesno{52},
pp. \bfpage{609}--\blpage{631}
(\byear{2023}).
\bcomment{Elsevier}
\end{bchapter}
\endbibitem

%%% 29
\bibitem[\protect\citeauthoryear{{Norwegian Public Road
  Administration}}{2022}]{VegLCA}
\begin{botherref}
\oauthor{\bsnm{{Norwegian Public Road Administration}}}:
VegLCA.
Online.
v5.11B
(2022)
\end{botherref}
\endbibitem

%%% 30
\bibitem[\protect\citeauthoryear{NPRA}{2025}]{N400}
\begin{bbook}
\bauthor{\bsnm{NPRA}}:
\bbtitle{N400: Bruprosjektering}.
\bpublisher{{The Norwegian Public Roads Administration}},
\blocation{Norway}
(\byear{2025})
\end{bbook}
\endbibitem

%%% 31
\bibitem[\protect\citeauthoryear{Nocedal and
  Wright}{1999}]{nocedal1999numerical}
\begin{bbook}
\bauthor{\bsnm{Nocedal}, \binits{J.}},
\bauthor{\bsnm{Wright}, \binits{S.J.}}:
\bbtitle{Numerical Optimization}.
\bpublisher{Springer},
\blocation{New York, NY}
(\byear{1999})
\end{bbook}
\endbibitem

%%% 32
\bibitem[\protect\citeauthoryear{Paul-Dubois-Taine and
  Amsallem}{2015}]{paul2015adaptive}
\begin{barticle}
\bauthor{\bsnm{Paul-Dubois-Taine}, \binits{A.}},
\bauthor{\bsnm{Amsallem}, \binits{D.}}:
\batitle{An adaptive and efficient greedy procedure for the optimal training of
  parametric reduced-order models}.
\bjtitle{International Journal for Numerical Methods in Engineering}
\bvolume{102}(\bissue{5}),
\bfpage{1262}--\blpage{1292}
(\byear{2015})
\end{barticle}
\endbibitem

%%% 33
\bibitem[\protect\citeauthoryear{Penad{\'e}s-Pl{\`a}
  et~al.}{2019}]{penades2019accelerated}
\begin{barticle}
\bauthor{\bsnm{Penad{\'e}s-Pl{\`a}}, \binits{V.}},
\bauthor{\bsnm{Garc{\'\i}a-Segura}, \binits{T.}},
\bauthor{\bsnm{Yepes}, \binits{V.}}:
\batitle{Accelerated optimization method for low-embodied energy concrete
  box-girder bridge design}.
\bjtitle{Engineering Structures}
\bvolume{179},
\bfpage{556}--\blpage{565}
(\byear{2019})
\end{barticle}
\endbibitem

%%% 34
\bibitem[\protect\citeauthoryear{Peherstorfer and
  Willcox}{2015}]{peherstorfer2015dynamic}
\begin{barticle}
\bauthor{\bsnm{Peherstorfer}, \binits{B.}},
\bauthor{\bsnm{Willcox}, \binits{K.}}:
\batitle{Dynamic data-driven reduced-order models}.
\bjtitle{Computer Methods in Applied Mechanics and Engineering}
\bvolume{291},
\bfpage{21}--\blpage{41}
(\byear{2015})
\end{barticle}
\endbibitem

%%% 35
\bibitem[\protect\citeauthoryear{Rocha et~al.}{2023}]{rocha2023reduced}
\begin{barticle}
\bauthor{\bsnm{Rocha}, \binits{P.R.B.}},
\bauthor{\bsnm{Sousa~Almeida}, \binits{J.L.}},
\bauthor{\bsnm{Paula~Gomes}, \binits{M.S.}},
\bauthor{\bsnm{Junior}, \binits{A.C.N.}}:
\batitle{Reduced-order modeling of the two-dimensional rayleigh--b{\'e}nard
  convection flow through a non-intrusive operator inference}.
\bjtitle{Engineering Applications of Artificial Intelligence}
\bvolume{126},
\bfpage{106923}
(\byear{2023})
\end{barticle}
\endbibitem

%%% 36
\bibitem[\protect\citeauthoryear{R{\o}stum
  et~al.}{2025}]{rostum2025constrained}
\begin{barticle}
\bauthor{\bsnm{R{\o}stum}, \binits{H.}},
\bauthor{\bsnm{Gros}, \binits{S.}},
\bauthor{\bsnm{Aas-Jakobsen}, \binits{K.}}:
\batitle{Constrained bayesian optimization for engineering bridge design}.
\bjtitle{Structural and Multidisciplinary Optimization}
\bvolume{68}(\bissue{1}),
\bfpage{1}--\blpage{14}
(\byear{2025})
\end{barticle}
\endbibitem

%%% 37
\bibitem[\protect\citeauthoryear{{Standard Norge}}{2021}]{EC2-1-1}
\begin{bbook}
\bauthor{\bsnm{{Standard Norge}}}:
\bbtitle{Eurocode 2: Design of Concrete Structures, Part 1-1: General Rules and
  Rules for Buildings}.
\bpublisher{Standard Norge},
\blocation{Oslo, Norway}
(\byear{2021}).
\bcomment{EN 1992-1-1:2004+A1+NA}
\end{bbook}
\endbibitem

%%% 38
\bibitem[\protect\citeauthoryear{Saves et~al.}{2024}]{saves2024gaussian}
\begin{bchapter}
\bauthor{\bsnm{Saves}, \binits{P.}},
\bauthor{\bsnm{Van}, \binits{E.N.}},
\bauthor{\bsnm{Bartoli}, \binits{N.}},
\bauthor{\bsnm{Lef{\`e}bvre}, \binits{T.}},
\bauthor{\bsnm{Diouane}, \binits{Y.}},
\bauthor{\bsnm{Morlier}, \binits{J.}}:
\bctitle{Gaussian process for bayesian optimization with mixed hierarchical
  variables: Application to electric-hybrid aircraft eco-design}.
In: \bbtitle{Workshop on Bayesian Optimization \& Related Topics}
(\byear{2024})
\end{bchapter}
\endbibitem

%%% 39
\bibitem[\protect\citeauthoryear{Tfaily et~al.}{2024}]{tfaily2024bayesian}
\begin{barticle}
\bauthor{\bsnm{Tfaily}, \binits{A.}},
\bauthor{\bsnm{Diouane}, \binits{Y.}},
\bauthor{\bsnm{Bartoli}, \binits{N.}},
\bauthor{\bsnm{Kokkolaras}, \binits{M.}}:
\batitle{Bayesian optimization with hidden constraints for aircraft design}.
\bjtitle{Structural and Multidisciplinary Optimization}
\bvolume{67}(\bissue{7}),
\bfpage{123}
(\byear{2024})
\end{barticle}
\endbibitem

%%% 40
\bibitem[\protect\citeauthoryear{{United Nations Environment Programme} and
  {Yale Center for Ecosystems + Architecture}}{2023}]{UN2023_Climate}
\begin{botherref}
\oauthor{\bsnm{{United Nations Environment Programme}}},
\oauthor{\bsnm{{Yale Center for Ecosystems + Architecture}}}:
Building Materials and the Climate: Constructing a New Future
(2023).
\url{https://wedocs.unep.org/20.500.11822/43293}
\end{botherref}
\endbibitem

%%% 41
\bibitem[\protect\citeauthoryear{Virtanen et~al.}{2020}]{scipy}
\begin{barticle}
\bauthor{\bsnm{Virtanen}, \binits{P.}},
\bauthor{\bsnm{Gommers}, \binits{R.}},
\bauthor{\bsnm{Oliphant}, \binits{T.E.}},
\bauthor{\bsnm{Haberland}, \binits{M.}},
\bauthor{\bsnm{Reddy}, \binits{T.}},
\bauthor{\bsnm{Cournapeau}, \binits{D.}},
\bauthor{\bsnm{Burovski}, \binits{E.}},
\bauthor{\bsnm{Peterson}, \binits{P.}},
\bauthor{\bsnm{Weckesser}, \binits{W.}},
\bauthor{\bsnm{Bright}, \binits{J.}}, \betal:
\batitle{Scipy 1.0: fundamental algorithms for scientific computing in python}.
\bjtitle{Nature methods}
\bvolume{17}(\bissue{3}),
\bfpage{261}--\blpage{272}
(\byear{2020})
\end{barticle}
\endbibitem

%%% 42
\bibitem[\protect\citeauthoryear{Williams and Rasmussen}{2006}]{Book_GPForML}
\begin{bbook}
\bauthor{\bsnm{Williams}, \binits{C.K.}},
\bauthor{\bsnm{Rasmussen}, \binits{C.E.}}:
\bbtitle{Gaussian Processes for Machine Learning}.
\bpublisher{MIT press},
\blocation{Cambridge, MA}
(\byear{2006})
\end{bbook}
\endbibitem

%%% 43
\bibitem[\protect\citeauthoryear{Yang et~al.}{2019}]{yang2019efficient}
\begin{barticle}
\bauthor{\bsnm{Yang}, \binits{K.}},
\bauthor{\bsnm{Emmerich}, \binits{M.}},
\bauthor{\bsnm{Deutz}, \binits{A.}},
\bauthor{\bsnm{B{\"a}ck}, \binits{T.}}:
\batitle{Efficient computation of expected hypervolume improvement using box
  decomposition algorithms}.
\bjtitle{Journal of Global Optimization}
\bvolume{75},
\bfpage{3}--\blpage{34}
(\byear{2019})
\end{barticle}
\endbibitem

%%% 44
\bibitem[\protect\citeauthoryear{Yepes~Llorente et~al.}{2024}]{yepes2024hybrid}
\begin{barticle}
\bauthor{\bsnm{Yepes~Llorente}, \binits{L.}},
\bauthor{\bsnm{Morlier}, \binits{J.}},
\bauthor{\bsnm{Sridhara}, \binits{S.}},
\bauthor{\bsnm{Suresh}, \binits{K.}}:
\batitle{A hybrid machine learning and evolutionary approach to material
  selection and design optimization for eco-friendly structures}.
\bjtitle{Structural and Multidisciplinary Optimization}
\bvolume{67}(\bissue{5}),
\bfpage{69}
(\byear{2024})
\end{barticle}
\endbibitem

%%% 45
\bibitem[\protect\citeauthoryear{Zhao}{2021}]{zhao2021reduced}
\begin{barticle}
\bauthor{\bsnm{Zhao}, \binits{H.}}:
\batitle{A reduced order model based on machine learning for numerical
  analysis: An application to geomechanics}.
\bjtitle{Engineering Applications of Artificial Intelligence}
\bvolume{100},
\bfpage{104194}
(\byear{2021})
\end{barticle}
\endbibitem

\end{thebibliography}

\end{document}